\documentclass[final]{IEEEtran}
\usepackage[table,xcdraw]{xcolor}
\usepackage{graphicx,subfig,xcolor,setspace,amsthm,amsmath,cite,booktabs,epstopdf, gensymb, dblfloatfix,blindtext,amsthm, booktabs, arydshln}
\usepackage{ textcomp }
\usepackage[hyphens]{url}
\usepackage{bm}
\usepackage{algorithmic}
\usepackage{upgreek,acro}
\usepackage{bbm}
\usepackage[ruled,norelsize]{algorithm2e}
\usepackage{setspace}
\usepackage{microtype}
\usepackage[final]{changes}
\usepackage{array}
\usepackage[binary-units,detect-weight=true, detect-family=true]{siunitx}
\DeclareSIUnit \bitspersecond {bps}

\usepackage{amsmath,amssymb}
\newcommand{\bbC}{{\mathbb{C}}}

\newcommand{\bbE}{{\mathbb{E}}}

\newcommand{\bbR}{{\mathbb{R}}}

\newcommand{\ba}{{\mathbf{a}}}

\newcommand{\bc}{{\mathbf{c}}}

\newcommand{\bi}{{\mathbf{i}}}
\newcommand{\bj}{{\mathbf{j}}}

\newcommand{\bn}{{\mathbf{n}}}

\newcommand{\bp}{{\mathbf{p}}}

\newcommand{\bs}{{\mathbf{s}}}

\newcommand{\bx}{{\mathbf{x}}}
\newcommand{\by}{{\mathbf{y}}}
\newcommand{\bz}{{\mathbf{z}}}

\newcommand{\bA}{{\mathbf{A}}}

\newcommand{\bC}{{\mathbf{C}}}

\newcommand{\bF}{{\mathbf{F}}}

\newcommand{\bH}{{\mathbf{H}}}
\newcommand{\bI}{{\mathbf{I}}}

\newcommand{\bK}{{\mathbf{K}}}
\newcommand{\bL}{{\mathbf{L}}}

\newcommand{\bN}{{\mathbf{N}}}

\newcommand{\bP}{{\mathbf{P}}}
\newcommand{\bQ}{{\mathbf{Q}}}
\newcommand{\bR}{{\mathbf{R}}}
\newcommand{\bS}{{\mathbf{S}}}

\newcommand{\bV}{{\mathbf{V}}}
\newcommand{\bW}{{\mathbf{W}}}
\newcommand{\bX}{{\mathbf{X}}}
\newcommand{\bY}{{\mathbf{Y}}}
\newcommand{\bZ}{{\mathbf{Z}}}
\newcommand{\rma}{{\mathrm{a}}}

\newcommand{\rmc}{{\mathrm{c}}}
\newcommand{\rmd}{{\mathrm{d}}}

\newcommand{\rmk}{{\mathrm{k}}}

\newcommand{\rmp}{{\mathrm{p}}}

\newcommand{\rmr}{{\mathrm{r}}}
\newcommand{\rms}{{\mathrm{s}}}
\newcommand{\rmt}{{\mathrm{t}}}

\newcommand{\rmx}{{\mathrm{x}}}
\newcommand{\rmy}{{\mathrm{y}}}

\newcommand{\rmA}{{\mathrm{A}}}
\newcommand{\rmB}{{\mathrm{B}}}

\newcommand{\rmD}{{\mathrm{D}}}

\newcommand{\rmF}{{\mathrm{F}}}

\newcommand{\cI}{\mathcal{I}}
\newcommand{\cJ}{\mathcal{J}}

\newcommand{\cN}{\mathcal{N}}
\newcommand{\cO}{\mathcal{O}}

\newcommand{\cS}{\mathcal{S}}

\newcommand{\sfT}{\mathsf{T}}

\newcommand{\btheta}{\boldsymbol{\theta}}

\newcommand{\bphi}{\boldsymbol{\phi}}

\newcommand{\bUpsilon}{\boldsymbol{\Upsilon}}
\newcommand{\bPhi}{\boldsymbol{\Phi}}
\newcommand{\bPsi}{\boldsymbol{\Psi}}

\makeatletter
\def\munderbar#1{\underline{\sbox\tw@{$#1$}\dp\tw@\z@\box\tw@}}
\makeatother

\DeclareAcronym{3GPP}{
	short=3GPP,
	long=3rd Generation Partnership Project
}
\DeclareAcronym{ADC}{
	short=ADC,
	long=analog-to-digital converter
}
\DeclareAcronym{AMP}{
	short=AMP,
	long=approximate message passing
}
\DeclareAcronym{AoA}{
	short=AoA,
	long=angle-of-arrival
}
\DeclareAcronym{AoD}{
	short=DoD,
	long=angle-of-departure
}
\DeclareAcronym{APS}{
	short=APS,
	long=azimuth power spectrum
}
\DeclareAcronym{AV}{
	short=AV,
	long=autonomous vehicle
}
\DeclareAcronym{BS}{
	short=BS,
	long=base station
}
\DeclareAcronym{BSM}{
	short=BSM,
	long=basic safety message
}
\DeclareAcronym{CDF}{
	short=CDF,
	long=cumulative distribution function
}
\DeclareAcronym{CNN}{
	short=CNN,
	long=convolutional neural network
}
\DeclareAcronym{CP}{
	short=CP,
	long=cyclic-prefix
}

\DeclareAcronym{CS}{
	short=CS,
	long=compressed sensing
}
\DeclareAcronym{CSI}{
	short=CSI,
	long=channel state information
}

\DeclareAcronym{DFT}{
	short=DFT,
	long=discrete Fourier transform
}
\DeclareAcronym{DL}{
	short=DL,
	long=deep learning
}

\DeclareAcronym{DNN}{
	short=DNN,
	long=deep neural network
}
\DeclareAcronym{DoA}{
	short=DoA,
	long=direction-of-arrival
}
\DeclareAcronym{DoD}{
	short=DoD,
	long=direction-of-departure
}
\DeclareAcronym{DSRC}{
	short=DSRC,
	long=dedicated short-range communication
}
\DeclareAcronym{EM}{
	short=EM,
	long=expectation maximization
}
\DeclareAcronym{FC}{
	short=FC,
	long=fully connected
}

\DeclareAcronym{FDD}{
	short=FDD,
	long=frequency division duplex
}
\DeclareAcronym{FMCW}{
	short=FMCW,
	long=frequency modulated continuous wave
}
\DeclareAcronym{FoV}{
	short=FoV,
	long=field-of-view
}
\DeclareAcronym{GNSS}{
	short=GNSS,
	long=global navigation satellite system
}
\DeclareAcronym{GPS}{
	short=GPS,
	long=global positioning system
}
\DeclareAcronym{IoT}{
	short=IoT,
	long=internet of things
}
\DeclareAcronym{IMU}{
	short=IMU,
	long=inertial measurement unit 
}
\DeclareAcronym{KL}{
	short=KL,
	long=Kullback–Leibler
}
\DeclareAcronym{LIDAR}{
	short=LIDAR,
	long=light detection and ranging
}
\DeclareAcronym{LOS}{
	short=LOS,
	long=line-of-sight
}
\DeclareAcronym{LPF}{
	short=LPF,
	long=low pass filter
}
\DeclareAcronym{LTE}{
	short=LTE,
	long=long term evolution
}
\DeclareAcronym{LS}{
	short=LS,
	long=least squares
}
\DeclareAcronym{LSTM}{
	short=LSTM,
	long=long short-term memory
}
\DeclareAcronym{mmWave}
{
	short = mmWave, 
	long = millimeter wave
}
\DeclareAcronym{MOMP}{
	short=MOMP,
	long=multidimensional orthogonal matching pursuit
}

\DeclareAcronym{MIMO}{
	short=MIMO,
	long=multiple-input multiple-output
}
\DeclareAcronym{MLE}{
	short=MLE,
	long=maximum likelihood estimation
}
\DeclareAcronym{MRR}{
	short=MRR,
	long=medium range radar
}
\DeclareAcronym{NLOS}{
	short=NLOS,
	long=non-line-of-sight
}

\DeclareAcronym{NLP}{
	short=NLP,
	long=natural language processing
}

\DeclareAcronym{NR}{
	short=NR,
	long=new radio
}
\DeclareAcronym{OFDM}{
	short=OFDM,
	long=orthogonal frequency-division multiplexing
}
\DeclareAcronym{OMP}{
	short=OMP,
	long=orthogonal matching pursuit
}
\DeclareAcronym{PDP}{
	short=PDP,
	long=power delay profiles
}
\DeclareAcronym{ppm}{
	short=ppm,
	long=parts-per-million
}

\DeclareAcronym{RMS}{
	short=RMS,
	long=root-mean-square
}
\DeclareAcronym{RPE}{
	short=RPE,
	long=relative precoding efficiency
}
\DeclareAcronym{RSU}{
	short=RSU,
	long=roadside unit
}
\DeclareAcronym{RTT}{
	short=RTT,
	long=round trip time
}
\DeclareAcronym{RX}{
	short=RX,
	long=receiver
}

\DeclareAcronym{RSRP}{
	short=RSRP,
	long=reference signal received power
 }

\DeclareAcronym{SNR}{
	short=SNR,
	long=signal-to-noise ratio
}
\DeclareAcronym{SLAM}{
	short=SLAM,
	long=simultaneous localization and mapping
}
\DeclareAcronym{SBL}{
	short=SBL,
	long=sparse Bayesian learning
}
\DeclareAcronym{SOMP}{
	short=SOMP,
	long=simultaneous orthogonal matching pursuit 
}

\DeclareAcronym{TCN}{
	short=TCN,
	long=temporal convolutional network
}
\DeclareAcronym{TX}{
	short=TX,
	long=transmitter
}

\DeclareAcronym{TDoA}{
	short=TDoA,
	long=time-difference-of-arrival
}

\DeclareAcronym{ToA}{
	short=ToA,
	long=time-of-arrival
}

\DeclareAcronym{UL}{
	short=UL,
	long=uplink
}
\DeclareAcronym{ULA}{
	short=ULA,
	long=uniform linear array 
}
\DeclareAcronym{URA}{
	short=URA,
	long=uniform rectangular array 
}
\DeclareAcronym{V2I}{
	short=V2I,
	long=vehicle-to-infrastructure
}
\DeclareAcronym{V2V}{
	short=V2V,
	long=vehicle-to-vehicle
}
\DeclareAcronym{V2X}{
	short=V2X,
	long=vehicle-to-everything
}
\DeclareAcronym{VRU}{
	short=VRU,
	long=vulnerable road user
}

\newcommand{\deh}[1]{\hspace{#1 mm}}

\newcommand{\sprl}{\shortparallel}

\definecolor{yRed}{HTML}{fc0600}
\definecolor{lightgray}{HTML}{b8b8aa}
\definecolor{cyedN}{HTML}{25a244}

\begin{document}
	
	\title{Learning to Localize with Attention: from Sparse mmWave Channel Estimates from a Single BS to High Accuracy 3D Location}
	\author{Yun Chen, {\it Student Member, IEEE}, Nuria Gonz\'alez-Prelcic, {\it Senior Member, IEEE}, Takayuki Shimizu and Chinmay Mahabal
		\thanks{A preliminary version of this work, which does not consider a location refinement stage was presented in \cite{SAM2022}. This material is based upon work partially supported by the National Science Foundation under grant no. 2433782  and is supported in part by funds from the federal agency and industry partners as specified in the Resilient \& Intelligent NextG Systems (RINGS) program and by a gift from Toyota Motor North America.}
		\thanks{Y. Chen and N. Gonz\'alez-Prelcic are with the Department of Electrical and Computer Engineering, University of California, San Diego, CA 92161, USA (e-mail:\{yuc216, ngprelcic\}@ucsd.edu).}
		\thanks{T. Shimizu and C. Mahabal are with Toyota Motor North America, Mountain View, CA 94043 USA (e-mail: \{takayuki.shimizu, chinmay.mahabal\}@toyota.com).}
	}
	%\markboth{For submission to IEEE Journal on Selected Areas in Communication}{}
	
	\maketitle
	
	%%%%%%%%%%%%%%%%%%%%%%%%%%%%%%%%%%%%%%%%%%%%%%%%%%%%%%%%%%%%%%%%%%%%%%%%%%%%%%%%%%%%%%%%%%%%%%%%%%%%%%%%%%%%%%
	\begin{abstract}
		One strategy to obtain user location information in a wireless network operating at \ac{mmWave} is based on the exploitation of the geometric relationships between the channel parameters and the user position. These relationships can be built from the \ac{LOS} path and first-order reflections, or purely first-order reflections, requiring high resolution channel estimates to ensure centimeter level accuracy. In this paper, we consider a mmWave \ac{MIMO} system employing a hybrid architecture, and develop a low complexity two-stage \ac{MOMP} algorithm suitable for accurate estimation of high dimensional channels. Then, a \ac{DNN} called \textit{PathNet} is designed to classify the order of the estimated channel paths, so that only the LOS path and first-order reflections are selected for localization. Next, a 3D localization strategy exploiting the geometry of the environment is developed to operate in both LOS and \ac{NLOS} conditions, while considering the unknown clock offset between the \ac{TX} and the \ac{RX}. Finally, a Transformer based network exploiting attention mechanisms called \textit{ChanFormer} is proposed to refine the initial position estimate obtained from geometric localization. {Simulation results obtained with realistic vehicular channels indicate that localization errors below $28$ cm can be achieved for $80$\% of the users when the LOS path is present, while sub-meter accuracy can be achieved for $55$\% of the users in NLOS conditions.}
	\end{abstract}
	
	%%%%%%%%%%%%%%%%%%%%%%%%%%%%%%%%%%%%%%%%%%%%%%%%%%%%%%%%%%%%%%%%%%%%%%%%%%%%%%%%%%%%%%%%%%%%%%%%%%%%%%%%%%%%%
	%%%%%%%%%%%%%%%%%%%%%%%%%%%%%%%%%%%%%%%%%%%%%%%%%%%%%%%%%%%%%%%%%%%%%%%%%%%%%%%%%%%%%%%%%%%%%%%%%%%%%%%%%%%%
	\section{Introduction}\label{sec:intro}
	%Introduction to the problem of joint localization and channel estimation to establish the mmWave link
	Wireless networks operating at \ac{mmWave} bands exploit large arrays and bandwidths, which lead to a high angle and delay resolvability when performing basic functions in the receiver such as channel parameter estimation, either for communication or localization purposes. {  In addition,   unlike at lower frequency bands --where the multipath is dense and becomes an interference for localization-- the sparsity of the mmWave channel makes it simpler to map the relevant channel paths to the geometry of the environment \cite{gonzalez2024integrated}. More specifically, the user location can be obtained from high resolution estimates of the multipath components of the channel between the user and a single \ac{BS}}, by exploiting the geometric relationships between the path parameters and the location of the scatterers, the \ac{BS} (assumed to be known), and the user \cite{Shahmansoori2018,gonzalez2024integrated}.
	This approach has the potential to become a cost-effective alternative for precise localization, required in many envisioned applications such as highly automated vehicles or robot automation in smart factories \cite{Wild2021}. In the vehicular setting, a \ac{CSI} based approach is robust to unfavorable weather or light conditions that may impact methods that exploit onboard automotive sensors, such as radars, \ac{LIDAR}, cameras, and \ac{IMU}s \cite{tao2022seqpolar, cwian2022gnss, kang2022lidar, schaefer2021long, liang2020scalable}. Moreover, it does not suffer from the low accuracy of  \ac{GNSS} in urban scenarios.
	Unfortunately, state-of-the-art solutions do not provide the required localization accuracy for some envisioned use cases --for example, an accuracy in the order of 0.1 m in vehicular settings \cite{Nokia5GNRPosWhitePaper2021}-- when evaluated in a realistic propagation environment with a practical mmWave MIMO architecture.
	\vspace*{-3mm}
	\subsection{Prior work}\label{PriorWork}
	Existing work on localization and channel estimation exploiting a snapshot from a single BS   \cite{Shahmansoori2018,gomez2023clock, Wymeersch2018, Talvitie2019,wen20205g, jiang2023low, Jiang2021, chen2022doppler, li2022joint, guan2023accurate,nazari2023mmwave,gong2023high,fascista2021downlink,xu2023joint} %Talvitie2019, Zhu2019}
can be based on two kinds of methods: 1) Two-stage approaches \cite{Shahmansoori2018, Wymeersch2018, Talvitie2019, gomez2023clock, chen2022doppler, Jiang2021, jiang2023low, wen20205g, li2022joint, guan2023accurate, nazari2023mmwave, gong2023high}, where the first stage focuses on channel parameter estimation, while the second stage has to solve an optimization problem to determine the user location from the channel path parameters, usually exploiting the geometry of the propagation environment. 2) Joint statistical approaches \cite{fascista2021downlink,xu2023joint}, which typically solve the optimization problems leveraging joint probability distributions of the parameters to be estimated. For the first category, the required accuracy of the channel estimation stage for precise localization is higher than for communication, since the estimated channel parameters are introduced into nonlinear geometric transformations very sensitive to estimation errors. Proposed channel estimation methods usually exploit the sparse nature of the \ac{mmWave} channel, and include on-grid approaches based on variations of \ac{OMP} such as SOMP, DCS-SOMP, or others \cite{Shahmansoori2018, Wymeersch2018, Talvitie2019, gomez2023clock}, off-grid strategies including subspace-based algorithms like multidimensional estimation via rotational invariance techniques (MD-ESPRIT) \cite{Jiang2021, jiang2023low, wen20205g, chen2022doppler} and atomic norm minimization \cite{li2022joint}, %Distributed Compressed Sensing-Simultaneous Orthogonal Matching Pursuit (DCS-SOMP) 
algorithms based on probability models such as the generalized turbo methodology  \cite{ guan2023accurate} or \ac{MLE} \cite{nazari2023mmwave} %and Bayesian learning \cite{Zhu2019},
to name a few. The user location can then be obtained by exploiting geometric relationships which involve path parameters and the known anchor node positions \cite{Shahmansoori2018, gomez2023clock, Jiang2021, jiang2023low, chen2022doppler, li2022joint, guan2023accurate}. %\tc{or by \ac{MLE} based on probability distributions, e.g., a factor graph is presented in \cite{Wymeersch2018} for deriving the likelihood functions; sampling methods are proposed in \cite{wen20205g, Talvitie2019}; and the Gaussian distribution assumption is made for the unknown parameters in \cite{nazari2023mmwave, gong2023high}. In addition, factoring in Doppler effects can aid localization \cite{chen2022doppler} considering moving vehicles.} 
In particular, when the user location has to be determined from the parameters of the channel between the user and a single BS, measurements of the \ac{DoA}, \ac{DoD}, and delay are required. If the channel is LOS, the user can be localized if these angular and delay measurements are available for the LoS path and at least one first-order reflection. In the NLOS scenario, the measurements for at least three first-order reflections are required \cite{SAM2022}. Methods of the second category consider a joint estimation of the channel and position parameters. The joint probability distributions of these parameters are exploited by methods such as \ac{MLE} \cite{fascista2021downlink} or \ac{EM} to reduce complexity \cite{xu2023joint}. The main limitation comes from the assumptions of specific joint distributions which may not hold for realistic channels.
% This type of method is stated to have higher accuracy.

To understand the limitations of prior work on joint localization and channel estimation at mmWave we focus first on reviewing previous work on channel estimation. Greedy strategies for mmWave channel estimation exploit the sparsity of the channel to obtain the parameters for every multipath component using a dictionary-based approach \cite{Wu2019, gurbuz2018sparse, mejri2020structured, SWOMP2018, Coma2018, zhang2023multi, Venugopal2017}. For example, a greedy approach based on a low-complexity \ac{OMP} algorithm that operates with a reduced dictionary constructed by exploiting statistical information of the scatters is proposed in  \cite{Wu2019}. A parameter perturbed \ac{OMP} algorithm is provided in \cite{gurbuz2018sparse}. Although a frequency-flat channel model is considered in \cite{Wu2019, gurbuz2018sparse}, other prior work \cite{mejri2020structured, SWOMP2018, Coma2018, zhang2023multi, Venugopal2017} offers dictionary-based solutions for frequency-selective mmWave channels. The simultaneous \ac{OMP} (SOMP) algorithm \cite{tropp2005simultaneous} is the core for the solutions developed in \cite{mejri2020structured, SWOMP2018 }. In \cite{mejri2020structured}, a joint subcarrier-block-based scheme is proposed using continually distributed angles of arrival and departure, while in \cite{SWOMP2018}, simultaneous weighted \ac{OMP} (SWOMP) is presented to account for the correlated noise after combining. In \cite{Coma2018}, the authors focus on compressive channel estimation on the uplink to configure precoders/combiners for the downlink based on channel reciprocity, and develop two algorithms for both purely digital and hybrid architectures. In \cite{zhang2023multi}, a multi-layer \ac{SBL} method for selectively increasing the angle resolution for channel estimation layer by layer helps to reduce the computational complexity and improves the performance. However, all these methods operate in the frequency domain, without estimating the path delays, which are required in any localization strategy that exploits the information about the channel between a user and a single \ac{BS}. A time domain channel estimation technique that accounts for the pulse shaping and filtering effect in the received signal, and can identify the path delays as well as angular parameters, was proposed in \cite{Venugopal2017}. However, it suffers from an extremely high complexity when operating with planar arrays and high resolution dictionaries.  To overcome this complexity limitation, an alternative approach called \ac{MOMP}, which also operates in the time domain estimating directions and delays, was recently proposed in \cite{MOMP2022}. The main idea behind MOMP is to operate with a multidimensional dictionary and perform the matching operation by independent tensor multiplications along each dimension, to later introduce a refinement stage based on alternating minimization.

Off-grid sparse recovery strategies were also developed in previous work to solve the channel estimation problem at mmWave \cite{you2022off, liu2022fast, yan2023off}. The method in \cite{you2022off} operates with a nonuniform grid, but it can be applied only to narrowband channels without filtering effects.  The key idea in \cite{liu2022fast} is to approximate a continuous infinite dictionary,  acquiring the channel parameters by solving a convex optimization problem, but again, it only applies to the unrealistic case of a narrowband channel without filtering effects. In \cite{yan2023off},  the authors assumed some specific distributions of the channel parameters, and a \ac{SBL}-based block \ac{EM} algorithm is proposed to perform Bayesian inference, assuming a MIMO-OTFS system, neglecting the filtering effects again, and focusing on time-varying channels. Other studies on off-grid mmWave channel estimation that focus on the narrowband case can be found in \cite{Qi2019,Anjinappa2019,Anjinappa2020,You2022TCOM}, but share the same limitation. An off-grid sparse recovery method is proposed in \cite{JRF2017CAMSAP} for frequency selective mmWave channels including the filtering effect, but it operates in the frequency domain and cannot be used to estimate the delay parameters required for single snapshot localization from a single \ac{BS}. Another category of off-grid methods exploits the ESPRIT algorithm \cite{wen2020tensor, zhang2021gridless}, but they also neglect the filtering effects in the channel model.
% However, off-grid methods prioritize the estimation of the dominant path/clusters, minimizing the normalized mean squared error (NMSE) between the estimated and actual channel, which cannot bring adequate number of estimated paths for localization.% \nuriacomment{We should include here a comment and citation to MOMP. It is a key element in the paper and we are skipping it}. \yuncommentY{Shall we include the comment below for MOMP here or after the DL method literature reviews?} \tc{To ensure the acquisition of a sufficient number of paths for localization, which relies on precise high-resolution 3D channel estimation in time domain while maintaining manageable complexity for modern computers, \ac{MOMP} algorithm \cite{MOMP2022} stands as the sole viable option, as it enables rearranging elements across multiple orthogonal dimensions and performing independent tensor multiplications along each dimension for large-scale tensors.
	
	Methods based on \ac{DL} have also been recently proposed to estimate the mmWave channel exploiting suitable datasets \cite{gao2023deep, liu2021sparsity, ma2020sparse}. Different network architectures have been proposed, including a 3D \ac{CNN} that approximates the sparse Bayesian learning process \cite{gao2023deep}, %\nuriacomment{is this paper about single shot positioning or position tracking?} \yuncommentY{It's about single-shot localization, and I agree the references for tracking should be removed.}, 
	a concatenated block architecture based on \ac{CNN} for extracting the channel coefficients \cite{liu2021sparsity}, and a \ac{FC} network for beamspace channel amplitude estimation and channel reconstruction \cite{ma2020sparse}. Strong limitations of all these DL methods are again that they operate in the frequency domain (i.e. the delays are not estimated), the discrete time channel model does not account for the pulse shaping and filtering stages at the receiver previous to analog-to-digital conversion, and the combining process is omitted in \cite{liu2021sparsity, ma2020sparse}, which results in the implicit assumption of using a fully digital architecture with high resolution converters, which is not feasible at mmWave.  
	
	Apart from using a \ac{DNN} for channel estimation to subsequently derive the user locations, \ac{DNN} can be directly applied to map channels to user locations based on channel fingerprinting, where channel characteristics such as \ac{RSRP} \cite{gao2023metaloc}, \ac{CSI} \cite{gao2023metaloc, salihu2022attention, lv2022deep}, beamformed fingerprints \cite{gante2020deep, wang2021deep}, angle-delay profiles \cite{wu2021learning}, etc., are leveraged. Networks based on \ac{CNN} architectures are proposed in \cite{lv2022deep, gante2020deep, wang2021deep, gao2023metaloc, wu2021learning}, where channel information can be structured as image-like inputs for convolutional layers to extract inherent features associated with user locations. While these methods require a stable environment with static channels, the work in \cite{salihu2022attention} considers dynamic environments and enables robust feature learning via attention schemes. However, these methods assume the availability of perfect channel information, without running into issues related to the computation complexity and inaccurate channel representations. In addition, localization accuracy is compromised when avoiding overfitting.

	In addition to the previously  discussed limitations of most of the previous work on the channel estimation strategy itself, specific work on model-based localization from a snapshot from a single BS  exploiting the channel parameters suffers from additional drawbacks: 1) an oversimplified communication system model, % which assumes all digital architectures \cite{Talvitie2019, Zhu2019}, or 
	which employs a limited number of antenna elements \cite{Shahmansoori2018, gomez2023clock, Talvitie2019, jiang2023low, chen2022doppler, li2022joint, guan2023accurate, nazari2023mmwave, gong2023high, xu2023joint, fascista2021downlink}, or neglects the filtering effects at the receiver \cite{Shahmansoori2018, gomez2023clock, Wymeersch2018, Talvitie2019, wen20205g, jiang2023low, Jiang2021, chen2022doppler, li2022joint, guan2023accurate, nazari2023mmwave,gong2023high, fascista2021downlink, xu2023joint}; 2) assumption of perfect TX-RX synchronization to exploit the \ac{ToA} \cite{Shahmansoori2018, Talvitie2019, wen20205g, jiang2023low, Jiang2021, li2022joint, guan2023accurate,gong2023high, fascista2021downlink, xu2023joint}; %, Talvitie2019, Zhu2019} 
3) artificially controlled evaluation settings which lead to simplistic and impractical channels, resulting in the lack of strategies to extract the \ac{LOS} and first-order \ac{NLOS} paths \cite{Shahmansoori2018, gomez2023clock, Wymeersch2018, Talvitie2019, wen20205g, jiang2023low, Jiang2021, chen2022doppler, li2022joint, guan2023accurate, nazari2023mmwave,gong2023high, fascista2021downlink, xu2023joint}; 4) high complexity of the 3D high resolution channel estimation process \cite{Shahmansoori2018,gomez2023clock, Talvitie2019}; %Talvitie2019, Zhu2019};
5) unsatisfactory localization accuracy--for example,  $\geq 10$ m  \cite{lv2022deep, gante2020deep}--when evaluated with realistic channels.
\vspace*{-3mm}
\subsection{Contributions}
% General introduction of our ideas (the whole pipeline).
In this paper,we propose a hybrid model/data-driven strategy for single shot joint localization and channel estimation. The data driven stage has been customized with data corresponding to vehicular channels,  but the strategy could be applied to any scenario by using the appropriate datasets.  Our approach begins by implementing a low complexity compressive channel estimation technique based on the MOMP algorithm \cite{Palacios2022Eusipco,MOMP2022}, which enables operation in realistic 3D environments. Then, a data driven method using \textit{PathNet} is employed to solve the path classification problem, identifying the necessary \ac{LOS} and first-order NLOS paths. A new position estimator, which can work in both LOS and NLOS channels with imperfect TX-RX synchronization, is then applied to convert the estimated parameters into the vehicle's 3D location. To further improve the localization accuracy, we introduce a novel strategy for position refinement -- a \ac{DNN} called \textit{ChanFormer} inspired by the \textit{Transformer} architecture \cite{Transformer2017}. %\nuriacomment{Add reference on self attention} 
It is used to analyze the estimated paths, evaluate the consistency between the estimated channel and the initial location estimate, and generate a probability distribution of the true position exploited to obtain a more precise location estimate.

% Contributions point by point:
The main contributions of the paper are as follows:
\begin{itemize}
\item We propose a realistic 3D mmWave channel model that includes the effects of the filtering stages at the receiver and the unknown clock offset between the TX and the RX, %which is not relevant for communication but 
which needs to be considered when the channel parameters are exploited for localization.
\item {We develop a two-stage MOMP channel estimation algorithm to reduce the complexity of the high resolution channel estimation process, turning computational burden from a product to a sum of terms.} It operates by jointly estimating first, for every path, the DoD in azimuth and elevation, the delay, and a parameter that contains a combination of the DoA information and the complex gain. An additional estimation stage is defined to retrieve the \ac{DoA} information.  
\item We build the lightweight yet effective \textit{PathNet} architecture for classifying the estimated channel paths. The training loss function is formulated to minimize the misclassification of high-order reflections as an \ac{LOS} or a first-order NLOS path. The network exhibits a strong generalization ability in new environments, achieving a classification accuracy of $99\%$. 
%\nuriacomment{Do we add the network parameters to Github?}
\item We develop a model-driven location estimator that exploits the channel geometry and can operate in both LOS and NLOS situations, as long as a sufficient number of paths are estimated. It provides sub-meter accuracy for more than  $85\%$ of the users in LOS vehicular channels and for $35\%$ of the users in purely NLOS channels.
\item We design \textit{ChanFormer}, a network that exploits the concept of ``attention" to evaluate which estimated paths are more credible and assess the likelihood of a given location being accurate. A mathematical formulation that models the likelihood based on the straight-line distance to the true location is proposed. \textit{ChanFormer} is intended to refine the location results obtained from the model-driven location estimator. 
%\yuncommentY{modified. } 
%{\item We analyze the error propagation throughout the system, examining how the two-stage MOMP estimation affects \textit{PathNet} classification, how \textit{PathNet} classification impacts geometric localization, and how \textit{PathNet } classification affects \textit{ChanFormer}'s performance. This analysis ensures our framework's robustness and its ability to tolerate errors introduced by each module. }
\item We generate a dataset containing realistic vehicular channels together with their associated vehicle positions generated by ray-tracing in an urban environment. All the simulations and evaluations of our algorithms are based on these channels, which are mostly composed of high-order NLOS paths. The dataset is available at \cite{Chen_Learning_to_Localize_2024} and can be used by the research community to evaluate any new solution to the joint localization and channel estimation problem in vehicular channels. {Simulation results show that 80\% of the users in LOS conditions experience localization errors below $28$ cm when exploiting our proposed strategy for localization, while sub-meter accuracy is achieved for 55\% of users in NLOS conditions.} %
\end{itemize}
Our overall scheme has been built upon our initial design in \cite{SAM2022}, completing all the details and derivations of the channel estimation strategy, extending the initial datasets for path classification, modifying the model-based initial localization strategy, adding the Transformer-based location refinement stage and including additional numerical experiments and comparisons with prior work.

%% Organization of the paper
The rest of the paper is structured as follows: Sec. \ref{sec:sys_model} describes the general \ac{V2I} communication setup, including the system model and the training strategy for joint channel estimation and localization. Sec. \ref{sec:HybridApproach} develops the different stages of our hybrid model/data-driven approach to joint localization and channel estimation. Then, Sec. \ref{sec:SimResults}, shows the numerical results of the experiments designed to evaluate the proposed strategy and the comparisons with previous work. Finally, Sec. \ref{conclu} concludes the paper, summarizing the main results and outlining future research directions.

\textbf{Notations:} Non-bold Italic letters $x$, $X$ are used for scalars; Bold lowercase $\bx$ is used for column vectors, and bold uppercase $\bX$ is used for matrices. $[\bx]_i$ and $[\bX]_{i,j}$, denote $i$-th entry of $\bx$ and entry at the $i$-th row and $j$-th column of $\bX$, respectively. $\bX^*$, $\bar{\bX}$ and $\bX^{\sfT}$ are the conjugate transpose, conjugate and transpose of $\bX$. $\|\bX \|_\rmF$ denotes the Frobenius norm of $\bX$. $[\bX,\bY]$ and $[\bX;\bY]$ are the horizontal and vertical concatenation of $\bX$ and $\bY$. $\cN(\bx,\bX)$ denotes a complex circularly symmetric Gaussian random vector with mean $\bx$ and covariance $\bX$. $\bI_N$ denotes a $N$-by-$N$ identity matrix. $\mathbb{N}$, $\mathbb{R}$, and $\mathbb{C}$ are the set of natural numbers, real numbers, and complex numbers, respectively. $\bbE[\cdot]$ denotes expectation. For mathematical calculations, $\bX\otimes\bY$, $\bX\odot\bY$, and $\bX\circ\bY$ are the Kronecker product, Hadamard product, and Khatri-Rao product of $\bX$ and $\bY$. $<\bx,\by>$ is the dot product of $\bx$ and $\by$.

\section{System model}\label{sec:sys_model}
%% Communication environment, illustrative figures,etc.
We consider a mmWave MIMO system where the users are active vehicles either communicating with the \ac{BS} or in initial access. The BS is equipped with a single \ac{URA}, and each vehicle is equipped with $4$ URAs facing front, back, right, and left, as suggested by the 3GPP methodology to simulate vehicular channels \cite{3GPPVehEnv}. %\begin{figure}[t!]
The URA at the BS  is equipped with $N_{\rmt}=N_{\rmt}^{\rmx}\times N_{\rmt}^{\rmy}$ antenna elements and is connected to $N_{\rmt}^{\rm RF}$ radio frequency (RF) chains, while each URA on the vehicle has $N_{\rmr}=N_{\rmr}^{\rmx}\times N_{\rmr}^{\rmy}$ antenna elements and is connected to $N_{\rmr}^{\rm RF}$ RF-chains. 
We focus on the downlink transmission during initial access, assuming hybrid analog-digital precoding and combining at both ends. We assume that $N_s$ data streams are transmitted, with $N_s\leq \min\{N_{\rmt}^{\rm RF}, N_{\rmr}^{\rm RF}\}$. The hybrid precoder is defined as ${\bf F}=\bF_{\rm RF}\bF_{\rm BB}\in\mathbb{C}^{N_{\rmt}\times N_s}$, and the hybrid combiner is ${\bW}=\bW_{\rm RF}\bW_{\rm BB}\in\mathbb{C}^{N_{\rmr}\times N_s}$, where the subscript RF stands for the analog counterpart of the precoder/combiner and BB for the digital one. We consider a fully connected phase shifting network \cite{mendez2016hybrid}. 

To develop the 3D channel model we define $\theta^{\rmx}$ and $\theta^{\rmy}$ as the \ac{DoA}  in azimuth and elevation, while $\phi^{\rmx}$ and $\phi^{\rmy}$ represent the \ac{DoD} also in both dimensions. Note that the azimuth and elevation angles are in the range of $[0, \pi)$ and $[-\frac{\pi}{2}, \frac{\pi}{2})$, respectively. The unitary vectors for the \ac{DoA} and \ac{DoD} are given by ${\boldsymbol \theta}=[\cos\theta^{\rmy}\cos\theta^{\rmx}, \cos\theta^{\rmy}\sin\theta^{\rmx}, \sin\theta^{\rmy}]^{\sfT}$, and ${\boldsymbol \phi}=[\cos\phi^{\rmy}\cos\phi^{\rmx}, \cos\phi^{\rmy}\sin\phi^{\rmx}, \sin\phi^{\rmy}]^{\sfT}$. Assuming the arrays are placed in the $yz$-plane with a half-wavelength element spacing, the array response at the \ac{RX} $\ba(\btheta)$ can be formulated where:
\begin{align}
\left[\ba(\btheta)\right]_{(n_\rmr^\rmx\deh{-0.5}-\deh{-0.5}1)N_\rmr^\rmy\deh{-0.5}+\deh{-0.2}n_\rmr^\rmy}=e^{-j\pi\left((n_\rmr^\rmx\deh{-0.5}-\deh{-0.5}1)\cos\theta^\rmy\sin\theta^\rmx+(n_\rmr^\rmy\deh{-0.5}-\deh{-0.5}1)\sin\theta^\rmy\right)}, 
\end{align}
which can be represented as the Kronecker product of two vectors as $\ba(\btheta)=\ba(\btheta^\sprl)\otimes\ba(\btheta^\bot)$ for later multidimensional operations, where $\left[\ba(\btheta^\sprl)\right]_{n}=e^{-j\pi(n-1)\cos\theta^\rmy\sin\theta^\rmx}$, and $\left[\ba(\btheta^\bot)\right]_{n}=e^{-j\pi(n-1)\sin\theta^\rmy}$.
%The array responses at the TX and RX are $\ba_{\rmt}({\boldsymbol\phi})=\ba_{\rmt}(\phi^{\rmx})\otimes \ba_{\rmt}(\phi^{\rmy})$ and $\ba_{\rmr}({\boldsymbol\theta})=\ba_{\rmr}(\theta^{\rmx})\otimes \ba_{\rmr}(\theta^{\rmy})$. The steering vector $\ba_{\rmt}(\phi^{\rmx})\in \mathbb{C}^{N_{\rmt}^{\rmx}\times 1}$  is defined as
% \begin{equation}
% \ba_{\rmt}(\phi^{\rmx})=\left[1,e^{-j2\pi\vartheta^{\rmx}\sin(\phi^{\rmx})},...,e^{-j(N^{\rmx}_{\rmt}-1)2\pi\vartheta^{\rmx}\sin(\phi^{\rmx})}\right]^{\sfT},
% \end{equation}
% where $\vartheta^{\rmx}$ is the inter-element spacing in the $x$-dimension normalized by the wavelength.
Similar definitions can be built for $\ba(\bphi)$ as $\ba(\bphi)=\ba(\bphi^\sprl)\otimes\ba(\bphi^\bot)$.  The effective discrete time baseband channel is seen through the RF front end, so the effects of the filtering stages before analog-to-digital conversion should be included in the channel model. We represent the overall response of the filtering stages by the function $f_{\rmp}$. The channel matrix for the  $n$-th delay tap is $\bH_n\in\mathbb{C}^{N_{\rmr}\times N_{\rmt}},\ n=0,...,N_{\rmd}-1$, which can be written as
\begin{equation}
\bH_n=\sum\limits_{\ell=1}^{L}\alpha_\ell f_{\rmp}\left(nT_s-(t_{\ell}-t_0)\right)\ba_{\rmr}({\boldsymbol\theta}_{\ell})\ba_{\rmt}({\boldsymbol\phi}_{\ell})^*,
\end{equation}
where $\alpha_{\ell}$ and $t_{\ell}$ are the complex gain and the \ac{ToA} of the $\ell$-th path, $T_s$ is the  sampling period, and $t_0$ is the unknown clock offset. Since the channel estimation algorithm in the mmWave receiver will provide an estimate of the relative delay $\tau_{\ell}=t_{\ell}-t_0$, $\ell=1,\ldots, L$, it is convenient to define the channel model as a function of $\tau_{\ell}$ instead of the absolute delays $t_{\ell}$.

During initial access, training signals are transmitted/received through several pairs of training precoders and combiners to sound the channel and localize the vehicle. We focus on the initial access stage, seeking to realize sub-meter vehicle localization accuracy as a byproduct of the link establishment. Further refinement of the location is possible by exploiting subsequent channel tracking stages, but it is out of the scope of this paper. 

Next, we build the model for the received signal during training. The $q$-th instance of the training sequence is a vector denoted as $\bs[q]\in \mathbb{C}^{N_s\times 1}, q=1,...,Q$, satisfying $\mathbb{E}[\bs[q]\bs[q]^*]=\frac{1}{N_s}\bI_{N_s}$. We consider a frequency selective MIMO channel with $N_{\rmd}$ delay taps. Assuming that the transmitted power is denoted as  $P_{\rmt}$, the $q$-th instance of the received signal can be written as
\begin{equation}\label{receive_sig}
\by[q]=\bW^*\sum\limits_{n=0}^{N_{\rmd}-1}\sqrt{P_{\rmt}}\bH_n\bF\bs[q-n]+\bW^*\bn[q],
\end{equation}
where $\bn[q]\sim \mathcal{N}({\bf 0},\sigma_{\bn}^2\bI_{N_\rmr})$ is additive white Gaussian noise. We compute the variance of the noise term as $\sigma^2_{\bn}=K_{\rmB}TB_c$, where $K_{\rmB}$ is the Boltzmann's constant, $T$ is the absolute temperature of the receiver, and $B_c$ is the system bandwidth.
Note that the noise after combining is no longer white, i.e. $\mathbb{E}[\bW^*\bn[q]\bn[q]^*\bW]=\sigma_{\bn}^2\mathbb{E}[\bW^*\bW]\neq \bI$. To whiten the receive signal in \eqref{receive_sig}, $\by[q]$ is multiplied by the inverse of a lower triangular matrix $\bL$ as $\breve{\by}[q]=\bL^{-1}\by[q]$, where $\bL$ is obtained from the Cholesky decomposition $\bW^*\bW=\bL\bL^*$.  Let $\breve{\bW}=\bL^{-1}\bW$ and $\breve{\bn}[q]=\bL^{-1}\bW^*\bn[q]$, then \eqref{receive_sig} can be rewritten as
\begin{equation}\label{whietened_receive_sig}
\breve{\by}[q]=\breve{\bW}^*\sum\limits_{n=0}^{N_{\rmd}-1}\sqrt{P_{\rmt}}\bH_n\bF\bs[q-n]+\breve{\bn}[q],
\end{equation}
where $\mathbb{E}[\breve{\bn}[q]\breve{\bn}[q]^*]= \sigma_{\bn}^2\bI$.  Let $\breve{\bY}=[\breve{\by} [1],...,\breve{\by} [Q]]\in \mathbb{C}^{N_s\times Q}$ be the matrix collecting the received samples for the different training frames, and $\breve{\bN} =[\breve{\bn} [1],...,\breve{\bn} [Q]]$ be the noise matrix. The whitened received signal matrix can be written as
\begin{equation}\label{receive_mat}
\breve{\bY} =\sqrt{P_\rmt}\breve{\bW}^* [\bH_0,...,\bH_{N_{\rmd}-1}]\left((\bI_{N_{\rmd}}\otimes\bF )\bS \right)+\breve{\bN},
\end{equation}
where 
\begin{equation}
\bS= 
\begin{bmatrix} \bs[1] & \bs[2] & \hdots & \bs[Q] \\ 
	{\bf 0} & \bs[1] & \hdots & \bs[Q-1] \\ 
	\vdots & \vdots & \ddots  & \vdots \\
	{\bf 0}  & {\bf 0}&  \hdots & \bs[Q-(N_{\rmd}-1)]
	
\end{bmatrix}.
\end{equation}
When using a set of $M_{\rmt}$ precoders $\{\bF_{m_{\rmt}}|m_{\rmt}=1,...,M_{\rmt}\}$ and a set of $M_{\rmr}$ combiners $\{\bW_{m_{\rmr}}|m_{\rmr}=1,...,M_{\rmr}\}$ for training, it is possible to write the expression of the received signal for a particular precoder/combiner pair $\breve{\bY}_{m_{\rmr},m_{\rmt}}$ from \eqref{receive_mat} by substituting $\bF$ by $\bF_{m_{\rmt}}$, $\breve{\bW}$ by $\breve{\bW}_{m_{\rmr}}$ and  $\breve{\bN}$ by $\breve{\bN}_{m_{\rmr}, m_{\rmt}}$.
% \nuriacomment{It is not clear at all how S changes with the particular training precoder/combiner pair. I thought we would send the same symbol sequence for training for the different combinations of the precoders and combiners, i.e., S should not change with mr/mt.} 
In the next sections, we develop the stages that process this received signal for channel estimation and precise positioning.

\section{Hybrid model/data driven approach for initial access and localization}\label{sec:HybridApproach}
\begin{figure*}[t!]
\centering
\includegraphics[width=\textwidth]{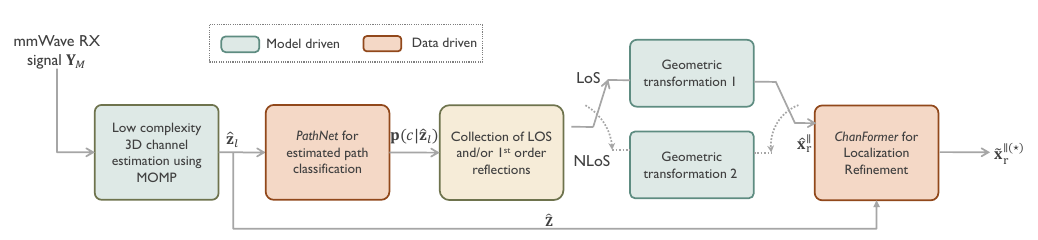}
\caption{Diagram of the joint initial access and localization system model. }
\label{Sys_model}
\vspace{-3mm}
\end{figure*}

The block diagram of our proposed joint initial access and 3D vehicle localization strategy is shown in Fig. \ref{Sys_model}. First, we collect in $\bY_M$ the mmWave received signals for $M$ different combinations of the training precoders and combiners.  Then, we employ a two-stage MOMP-based low complexity channel estimation technique to acquire $N_{\rm est}$ estimated paths $\hat{\bZ}=[\hat{\bz}_1,...,\hat{\bz}_{N_{\rm est}}]$, where each vector $\hat{\bz}_{\ell}$ contains: the magnitude of the estimated channel gain $|\alpha_{\ell}|$ 
% \nuriacomment{use $\ell$ everywhere for the channel path index}
, the relative delay $\tau_{\ell}=t_{\ell}-t_0$, the \ac{DoA} ${\boldsymbol\theta}_{\ell}$, and the \ac{DoD} ${\boldsymbol\phi}_{\ell}$. Then, every channel path is classified by \textit{PathNet}, a lightweight network which predicts the probability of $\hat{\bz}_{\ell}$ being a LOS component, a first-order reflection, or a high-order reflection, so that the \ac{LOS} and first-order reflections are later exploited for localization using the geometric relationships between the path parameters and the vehicle's position. Note that these relationships depend on the channel state (LOS or NLOS). Since the location estimates provided by this stage cannot guarantee sub-meter accuracy for most of the users, an additional data driven stage realized with \textit{ChanFormer}, a self-attention network,  is thus proposed for location refinement. To this aim, a set of tiles with a given size is built around the initial position estimate, and the output from \textit{ChanFormer} provides a probability map showing which tile contains the true location with the highest probability.   \textit{ChanFormer} analyzes the relationships among the estimated paths in $\hat{\bZ}$ and measures the congruence between the channel features and the initial location estimate $\hat{\bx}_{\rmr}^{\shortparallel}$. The input estimated channel features are extracted through self-attention in the encoder section of the network. These features are then decoded and matched to a more precise location estimate associated with the center of the highest probability tile. Though we use a square tile structure in this paper, the number and the shape of the tiles could be both customized to suit the specific environment and required accuracy.

% Sub-section: Communication system model

\subsection{Two-stage MOMP-based channel estimation}

\subsubsection{Channel estimation at mmWave exploiting sparsity and conventional OMP}

{   Prior work on compressive channel estimation at mmWave exploiting OMP (see for example \cite{Coma2018, SWOMP2018, venugopal2017channel}), leverages a representation of the channel in terms of a sparsifying dictionary $\bPsi$,  defined as a Kronecker product of several matrices built from the steering vectors at the transmitter and at the receiver evaluated on a grid for the \ac{DoD} and the \ac{DoA}, and an additional component to represent the delay domain. By definition, the Kronecker structure creates a dictionary with a size related to the product of the sizes of the matrices which represent the angular and delay domains. 
These sizes are also related to the array dimension and the required resolution of the dictionary. When operating at mmWave with large planar arrays, the size of the dictionary becomes too large, posing challenges in terms of memory and the number of operations required to solve the sparse recovery problem. Mathematically, the first step to define the sparse recovery problem consists of vectorizing the received signal matrix in (5). By exploiting the properties of the vectorization operator, we can obtain that 
\begin{equation}\label{vecY}
	\text{vec}(\breve{\bY}) = \bUpsilon\bPsi\bc + \text{vec}(\breve{\bN}), 
\end{equation}
where $\bUpsilon = \left((\bI_{N_\rmd}\otimes\bF)\sqrt{P_\rmt}\bS\right)^\sfT \otimes \breve{\bW}^*\in \bbC^{N_sQ\times N_\rmr N_\rmt N_\rmd}$ is the measurement matrix, $\bPsi\in\bbC^{N_\rmr N_\rmt N_\rmd\times N^{\rma}_\rmr N^{\rma}_\rmt N^{\rma}_\rmd}$ is the sparsifying dictionary, with $N^{\rma}_\rmr$, $N^{\rma}_\rmt$, and $N^{\rma}_\rmd$ depending on the required angular and delay resolutions, and $\bc\in \bbC^{N^{\rma}_\rmr N^{\rma}_\rmt N^{\rma}_\rmd\times 1}$ is the sparse vector representing the channel. The dictionary matrix $\bPsi$ is computed as [57]
\begin{align}
	\bPsi = \bA_\rmd\otimes(\overline{\bA}_\rmt\otimes \bA_\rmr)\in \bbC^{N_\rmr N_\rmt N_\rmd \times N_\rmr^\rma N_\rmt^\rma N_\rmd^\rma}, 
\end{align}
where $\bA_\rmd=\left[\bp(\ddot{t}_1), ...,\bp(\ddot{t}_{N^\rma_\rmd})\right]$ is the dictionary for the delay, being $\bp(t)=[f_\rmp(0\cdot T_s-t),\dots,\allowbreak f_\rmp((N_d-1)T_s-t)]^{\sfT}\in \bbR^{N_d\times 1}$ a sampled version of the function that models the filtering effects in the discrete time equivalent channel model in (2), and  $\{\ddot{t}_n| n=1,...,N_\rmd^\rma\}$ the grid points in the delay domain; $\bA_\rmt=\left[\ba(\ddot{\bphi}_1),...,\ba(\ddot{\bphi}_{N^\rma_\rmt})\right]\in\bbC^{N_\rmt\times N_\rmt^\rma}$ is the dictionary for the DoD considering the grid points $\{\ddot{\bphi}_n|n=1,...,N_\rmt^\rma\}$, and it  can be decomposed as $\bA_\rmt = \bA_\rmt^\rmx\otimes\bA_\rmt^\rmy$, where $[\bA^\rmx_\rmt]_{:,n} = \ba(\ddot{\bphi}^\sprl_n)$ and $[\bA^\rmy_\rmt]_{:,n} = \ba(\ddot{\bphi}^\bot_n)$, with $\ddot{\bphi}_n^\sprl$ and $\ddot{\bphi}_n^\bot$ the $n$-th selection of the grid values for the DoD in azimuth and elevation, respectively; finally, $\bA_\rmr=\left[\ba(\ddot{\btheta}_1),...,\ba(\ddot{\btheta}_{N_\rmr^\rma})\right]=\bA_\rmr^\rmx\otimes \bA_\rmr^\rmy\in \bbC^{N_\rmr\times N_\rmr^\rma}$ is the dictionary the DoA defined in a similar way as $\bA_\rmt$. 		Given these definitions, prior work (see for example [57]) estimates the sparse representation of the channel $\bc$ by exploiting OMP to solve the problem	\begin{align}\label{oriOpt}
	\min\limits_{\bc}\left\|\breve{\bY}-\bUpsilon\bPsi\bc\right\|^2, 
\end{align}
which has a complexity $\cO(N_{\rm est}N_sQN_\rmr N_\rmt N_\rmd N_\rmr^\rma N_\rmt^\rma N_\rmd^\rma)$. Given the practical values of the parameters that impact this complexity when operating with large antenna arrays and fine dictionary resolutions, the OMP algorithm could not be executed in a conventional server or a high end personal computer. To address this limitation, the recently proposed MOMP algorithm  \cite{Palacios2022Eusipco,MOMP2022} solves the associated sparse recovery problem by exploiting independent dictionaries for every sparse dimension instead of a single, very large dictionary, built as a Kronecker product of these independent dictionaries, as described next.
}

{
\subsubsection{MOMP based channel estimation}
The fundamental idea of MOMP is to rearrange elements in $\bUpsilon$ and $\bPsi$ into $N_{\rmD}$ orthogonal dimensions and execute tensor multiplications independently along each dimension. Considering the received signals $\breve{\bY}\in\bbC^{N_s\times Q}$, the algorithm starts by constructing $N_\rmD$ independent sparsifying dictionaries $\{{\bf \Psi}_{k}\in\mathbb{C}^{N_k^{\rm s}\times N_k^{\rm a}}\ |\ k=1,...,N_{\rmD}\}$, where $N_k^\rma$ is the number of atoms in the dictionary $\bPsi_k$, and $N_k^\rms$ is the size of each atom. Then,  the measurement tensor is defined as  ${\bf \Phi}\in\mathbb{C}^{N_s\times\otimes_{k=1}^{N_{\rm D}}N_k^{\rm s}}$,  %\nuriacomment{define missing parameters for the sensing matrix sizes} 
where $\otimes_{k=1}^{N_\rmD}N_k^\rms$ represents the tensor shape of $N_1^{\rms}\times N_2^{\rms}\times...\times N_\rmD^{\rms}$. The target of the algorithm is to solve the multidimensional matching pursuit problem in \eqref{eq:MOMP} to extract the sparse coefficients from the tensor ${\bf C}\in\mathbb{C}^{\otimes_{k=1}^{N_{\rm D}}N_k^{\rma}\times Q}$:
\begin{equation}\label{eq:MOMP}
	\min_{\bf C}\left(\left\|{\breve{\bY}}-\sum_{{\bf i}\in\mathcal{I}}\sum_{{\bf j}\in\mathcal{J}}[{\bf \Phi}]_{:, {\bf i}}\left(\prod_{k = 1}^{N_{\rm D}}[{\bf \Psi}_{k}]_{i_k, j_k}\right)[{\bf C}]_{{\bf j}, :}\right\|^2_{\rmF}\right),
\end{equation}
where $\mathcal{I} = \{{\bf i}=(i_1, \ldots, i_{N_{\rm D}})\in\mathbb{N}^{N_{\rm D}} | i_k \leq N_k^{\rm s}, \ \forall k \leq N_{\rm D}\}$, and $\mathcal{J} = \{{\bf j}=(j_1, \ldots, j_{N_{\rm D}})\in\mathbb{N}^{N_{\rm D}}|j_{d} \leq N_k^{\rm a}, \ \forall k \leq N_{\rm D}\}$ represent the multidimensional indices.}

The application of MOMP to joint localization and channel estimation in an indoor scenario was proposed in \cite{Palacios2022Eusipco,MOMP2022}, where all the additional details for the problem formulation and solution can be found, including a link to the algorithm implementation in GitHub (\url{https://github.com/WiSeCom-Lab/MOMP-core.git}).  In this work, $N_{\rmD}=5$ independent dictionaries are considered -- for the DoD in azimuth, DoD in elevation, delay domain, DoA in azimuth, and DoA in elevation, {namely, $\bPsi_1= \overline{\bA^\rmx_\rmt}$, $\bPsi_2=\overline{\bA^\rmy_\rmt}$, $\bPsi_3=\bA_\rmd$, $\bPsi_4=\bA_\rmr^\rmx$, and $\bPsi_5=\bA_\rmr^\rmy$.} { MOMP computes first the projections on these five different sparsifying dictionaries (which cover all possible angular domains and the delay component) independently, and exploits an alternating optimization strategy to converge to the same solution that conventional OMP would provide.  Although MOMP requires additional steps for initialization and iterative refinement, the complexity of each step is much lower than those in OMP, resulting in a much lower overall complexity. In particular,  the complexity is reduced to $\cO\left(N_{\rm est}N_sQN_{\rm iter}(\sum_{k=1}^{N_\rmD}N_k^\rma)(\prod_{k=1}^{N_\rmD}N_k^{\rms})\right)$ from the previous $\cO\left(N_{\rm est}N_sQ\prod_{k=1}^{N_\rmD}N_k^\rms N_k^\rma\right)$ when exploiting OMP, where 
$N_{\rm iter}$ is the number of iterations for refining the estimates associated with each dictionary. This advantage becomes particularly interesting when operating with  high resolution dictionaries, because both computational complexity and memory requirements are significantly reduced, since $\sum_{k=1}^{N_\rmD}N_{\rm iter}N_k^\rma \ll \prod_{k=1}^{N_\rmD}N_k^\rma$.}

\subsubsection{Two-stage MOMP enabling finer resolutions}
To further reduce memory requirements targeting the outdoor 3D localization problem, which considers large arrays and fine resolutions, we propose a modification of the MOMP-based channel estimation strategy in \cite{Palacios2022Eusipco,MOMP2022}. It comprises two stages:  1) estimating delays, DoDs, and a parameter that we define as equivalent gains, which include the combined effect of the path complex gains and the DoAs; this way we can apply MOMP for channel estimation with  $N_{\rmD}=3$ dictionaries instead of 5; and 2) estimating the DoAs from the equivalent gains. In the next paragraphs, we develop the estimators to implement this two-stage approach.

%\nuriacomment{I suggest that we start by defining the three dictionaries we are going to use, then defining the channel as a function of those dictionaries and the matrix to be estimated $\bC$ (see equation 32 in MOMP paper as an example of what I mean, but the expression will be different now);} 
{For stage 1), we start by constructing three independent sparsifying dictionaries $\bPsi_1$, $\bPsi_2$, and $\bPsi_3$ as defined before.}
% \begin{equation}\label{eq:Dictionaries}
% \begin{cases}
	% 	{\bf \Psi}_1 = [\bar{\bf a}_{\rmt}^{\rm x}(\ddot{\bphi}_1^{\sprl}), \cdots, \bar{\bf a}_{\rmt}^{\rm x}(\ddot{\phi}_{N_1^{\rm a}}^{\sprl})]\\
	% 	{\bf \Psi}_2 = [\bar{\bf a}_{\rmt}^{\rm y}(\ddot{\phi}_1^{\bot}), \cdots, \bar{\bf a}_{\rmt}^{\rm y}(\ddot{\phi}_{N_2^{\rm a}}^{\bot})]\\
	% 	{\bf \Psi}_3 = [{\bp}_{\rmd}(\ddot{t}_1), \cdots, {\bp_{\rmd}}(\ddot{t}_{N_3^{\rm a}})]
	% \end{cases},
% \end{equation}
%where ${\bf \Psi}_1$ and ${\bf \Psi}_2$ represent the departure angular domain in azimuth and elevation, and are built from the steering vectors at the transmitter evaluated on a grid of possible directions  $\{\ddot{\phi}_1^{\sprl}, \ldots, \ddot{\phi}_{N_1^{\rm a}}^{\sprl}\}$ and $\{\ddot{\phi}_1^{\bot}, \ldots, \ddot{\phi}_{N_2^{\rm a}}^{\bot}\}$; ${\bf \Psi}_3$ represents the delay domain $\{\ddot{t}_1, \ldots, \ddot{t}_{N_3^{\rm a}}\}$. 
%In the case, $N_1^\rms=N_\rmt^\rmx$, $N_2^\rms=N_\rmt^\rmy$, and $N_3^\rms=N_\rmd$.
Considering training with  $M_\rmr$ combiners, the effect of combiners and the arrival angular information are embedded into what we define as the equivalent gain tensor ${\bf C}\in\mathbb{C}^{N_{\rm 1}^{\rma}\times N_{\rm 2}^{\rma}\times N_{\rm 3}^{\rma}\times N_s M_{\rmr}}$, which is sparse, and can be written as
\begin{equation}\label{eq:Coefficients}
\vspace*{-3mm}
[{\bf C}]_{{\bf j}, :} = \left\lbrace\begin{array}{cl}
	{\boldsymbol \beta}_{\ell}^{\sfT} & \text{if }\begin{array}{c}
		\phi_{\ell}^{\rm x} = \ddot{\phi}_{j_1}^{\sprl}, \phi_{\ell}^{\rm y} = \ddot{\phi}_{j_2}^{\bot},\\
		t_{\ell}-t_0 = \ddot{t}_{j_3}
	\end{array};\\
	0 & \text{o.w.}
\end{array}\right. ,
\end{equation}
where $[{\boldsymbol \beta}_{\ell}]_{N_s (m_{\rmr}-1)+n_s}=\alpha_{\ell}[\breve{\bf W}_{m_{\rmr}}]^{*}_{:, n_s}{\bf a}_{\rmr}({\boldsymbol \theta}_{\ell})$.
%\nuriacomment{Now, from here,  we should be able to write the product $\breve{\bW}^* [\bH_0,...,\bH_{N_{\rmd}-1}]$ in (5) as a function of the three dictionaries and the tensor $\bC$ that you have just defined. This is the first step to be able to define the measurement tensor, which is the only component left in the MOMP optimization problem.} 
For a given training combiner  $\breve{\bW}_{m_\rmr}$, we define the part in \eqref{receive_mat} that contains the combiner and the channel matrices for different delays as the combined channel $\bH^{(m_\rmr)}= \breve{\bW}_{m_\rmr}^*\left[\bH_0,...,\bH_{N_\rmd-1}\right]$, with $\bH^{(m_\rmr)}\in\mathbb{C}^{N_s\times N_\rmd N_\rmt^\rmx N_\rmt^\rmy}$, which can also be represented by the multiplication of $\bPsi_k$ and $\bC$ as
\begin{align}
[\bH^{(m_\rmr)}&]_{n_s,(i_3-1)N_\rmt^\rmx N_\rmt^\rmy+(i_1-1)N_\rmt^\rmy+i_2} = \\ &\sum\limits_{\bj\in\cJ}\left(\prod_{k=1}^{3}[\bPsi_k]_{i_k,j_k}\right)[\bC]_{\bj,(m_\rmr-1)N_s+n_s}.
\end{align}
%\nuriacomment{ then we define the expression of the received signal as a function of the dictionaries, the tensor $\bC$ the precoder, the pilot and the noise, } 
Now \eqref{receive_mat} can be alternatively written as
\begin{align}
[&\breve{\bY}_{m_\rmr, m_\rmt}]_{n_s,q} =\\
&\sum\limits_{\bi\in \cI}[(\bI_{N_\rmd} \otimes\bF_{m_\rmt})\sqrt{P_\rmt}\bS]_{(i_3-1)N_\rmt^\rmx N_\rmt^\rmy+(i_1-1)N_\rmt^\rmy+i_2, q}  \\
& \cdot \sum\limits_{\bj\in\cJ}\left(\prod_{k=1}^{3}[\bPsi_k]_{i_k,j_k}\right)[\bC]_{\bj,(m_\rmr-1)N_s+n_s}+[\breve{\bN}]_{n_s,q}.
\end{align}
% \begin{align}
%  [\breve{\bY}_{m_\rmr, m_\rmt}]_{n_s,q} &=\\
%  & \sqrt{P_\rmt}\sum\limits_{\bi\in \cI}[{\bf F}_{m_{\rmt}}{\bf s}[q-(i_3-1)]]_{(i_1-1)N_{\rmt}^{\rmy}+i_2}  \\
% & \cdot \sum\limits_{\bj\in\cJ}\left(\prod_{k=1}^{3}[\bPsi_k]_{i_k,j_k}\right)[\bC]_{\bj,(m_\rmr-1)N_s+n_s} \\
% &+[\breve{\bN}]_{n_s,q}.
% \end{align}

%\nuriacomment{so we will be able to deduce the expression of the measurement tensor by comparing to equation 5. We can also extract the definitions of the different tensors by comparing to the general case in the MOMP paper, but every tensor has to be defined and justified where that definition comes from.} 
We can now derive the measurement tensor, which is the remaining key component for solving the MOMP problem. In \cite{Palacios2022Eusipco,MOMP2022}, the measurement tensor $\bPhi$ includes the effect of both precoder and combiner, however, it currently only contains the information of the precoder in our solution, with the combiner effects factored into the equivalent gain $\bC$ to be estimated. Hence, we define ${\boldsymbol\Phi}_{m_{\rmt}}\in \mathbb{C}^{Q\times N_{\rmt}^{\rmx}\times N_{\rmt}^{\rmy}\times N_{\rmd}}$ as the measurement tensor obtained with $\bF_{m_{\rmt}}$, 
%\nuriacomment{this cannot be right, since we say later that the effect of the combiner is in $\bC$. The measurement tensor cannot depend on the combiner index, it only depends on the precoder and the pilot} 
and the whole measurement tensor composed of ${\boldsymbol\Phi}_{m_{\rmt}}$, where $1\leq m_\rmt\leq M_\rmt$, {is ${\boldsymbol\Phi_M}\in \mathbb{C}^{QM_{\rmt} \times N_{\rmt}^{\rmx}\times N_{\rmt}^{\rmy}\times N_{\rmd}}$, where 
\begin{align}
	[{\bf \Phi}_M&]_{Q (m_{\rmt}-1)+q, {\bf i}}=\\&
	[(\bI_{N_\rmd}\otimes\bF_{m_\rmt})\bS]_{(i_3-1)N_\rmt^\rmx N_\rmt^\rmy+(i_1-1)N_\rmt^\rmy+i_2, q} = \\ &\ \ \ \ \left[{\bf F}_{m_{\rmt}}{\bf s}[q-(i_3-1)]\right]_{(i_1-1)N_{\rmt}^{\rmy}+i_2}.
	\end{align}}
	%\nuriacomment{Should not we use here the whitened combiner}. \yuncommentY{Yes, now all the whitened signals are using $\breve{\cdot}$.}
	
	Now the components of \eqref{eq:MOMP} are ready, where $\breve{\bY}_M$ is formed by collecting multiple observations using different pairs of $\bF_{m_{\rmt}}$ and $\bW_{m_{\rmr}}$:
	\begin{align}\label{observe_Y}
\breve{\bY}_M = \left[\begin{array}{ccc}
	\breve{\bf Y}^\sfT_{1, 1}& \cdots & \breve{\bf Y}^\sfT_{M_{\rmr}, 1}\\
	\vdots & \ddots & \vdots\\
	\breve{\bf Y}^\sfT_{1, M_{\rmt}} & \cdots & \breve{\bf Y}^\sfT_{M_{\rmr}, M_{\rmt}}\\
\end{array}\right]\in\mathbb{C}^{Q M_{\rmt}\times N_s M_{\rmr}}.
\end{align}
%Our particular MOMP problem to be solved is formulated as \eqref{eq:MOMP} with $\bO$ and $N_\rmD$ being replaced by $\breve{\bY}_M$ and $3$, respectively.
%\nuriacomment{For consistency with equation 5, we should use $\breve{\bY}_M$ in the equation above, since we work with the whitened version of the received signal.} 
% Our particular multidimensional matching pursuit problem to be solved can be written as:
% \begin{align}
% 	\min\limits_{\bC}\left(\left\|\breve{\bY}_M-\sum\limits_{\bi\in\cI}[\bPhi]_{:, \bi}\sum\limits_{\bj\in\cJ}\left(\prod_{k=1}^3[\bPsi_k]_{i_k,j_k}\right)[\bC]_{\bj,:} \right\|_\rmF^2\right).
% \end{align}
%\nuriacomment{Include equation of the optimization problem being solved as a function of the tensors just defined for clarity}. 
We employ the MOMP algorithm in \cite{MOMP2022} to solve this problem and obtain the estimated values of ${\bphi}_{\ell}$, $\tau_\ell$, and  ${\boldsymbol \beta}_{\ell}$, $\ell=1,\dots, L$.

For stage 2), to retrieve the DoA information from the non-zero coefficients of ${\bf C}$, i.e.  ${\boldsymbol \beta}_{\ell}$,  the main idea is to correlate the coefficients with angular dictionaries, so the DoA for the different paths can be obtained by finding the peaks of this correlation. 
%Let $\bPsi_4$ and $\bPsi_5$ be the dictionaries for the angle of arrival in azimuth and elevation, defined in a similar way to $\bPsi_1$ and $\bPsi_2$ in \eqref{eq:Dictionaries}, but considering the receive steering vector ${\ba}_{\rmr}$ instead of ${\ba}_{\rmt}$, and angular grids for the angles of arrival in azimuth and elevation, i.e., $\{\ddot{\theta}_1^{\sprl}, \ldots, \ddot{\theta}_{N_4^{\rm a}}^{\sprl}\}$ and $\{\ddot{\theta}_1^{\bot}, \ldots, \ddot{\theta}_{N_5^{\rm a}}^{\bot}\}$. 
Let $\bPsi_{\rmr}=\bPsi_4\otimes \bPsi_5$ and $\breve{\bW}_M = [\breve{\bW}_1, \cdots, \breve{\bW}_{M_{\rmr}}]$ (note that we remove the notation for measurement indices for simplicity), then ${\boldsymbol \beta}_{\ell}$ can be rewritten as ${\boldsymbol \beta}_{\ell} = \alpha_{\ell}{\breve{\bW}}_M^{* }{\ba}_{\rmr}({\boldsymbol \theta}_{\ell})$. Hence, assuming every entry of $\breve{\bW}_M$ is orthonormal, by multiplying $\boldsymbol\beta_{\ell}^*$, $\breve{\bf W}_M^{*}$, and the angular dictionary leads to
\begin{equation}
{\boldsymbol\beta}_{\ell}^* \breve{\bW}_M^*{\bf \Psi}_{\rmr}=\alpha_{\ell}{\ba}_{\rmr}({\boldsymbol \theta}_{\ell})^{*}\breve{\bW}_M\breve{\bW}^*_M{\bf \Psi}_{\rmr}=\alpha_{\ell} \ba ({\boldsymbol \theta}_{\ell})^*{\bf\Psi}_{\rmr},
\end{equation}
and the DoAs can now be retrieved as
\begin{equation}\label{GetDoA}
\hat{\boldsymbol \theta}_{\ell} = \arg\max_{\ddot{\boldsymbol \theta}} \hat{\boldsymbol \beta}_{\ell}^*\breve{\bf W}_M^*{\bf\Psi}_{\rmr}.
\end{equation}
Note that \eqref{GetDoA} can also be solved using MOMP by independent tensor multiplications in the arrival angular domain in azimuth and elevation, especially when $N_k^\rms$ and $N_k^\rma$ are large. 
{
This way, the computational complexity of our two-stage channel estimation algorithm is  $\cO\left(N_{\rm est}N_sQN_{\rm iter}(\sum_{k=1}^3N_k^\rma)(\prod_{k=1}^3N_k^\rms)\right)$,
which is lower than using the single stage MOMP for simultaneous estimation across five dimensions \cite{MOMP2022}. Table \ref{chanEstComplex} summarizes the computational complexity for the three channel estimation methods.} 
%cannot be handled by high-end servers when considering the dimensions of practical mmWave planar arrays and training sizes.
The channel parameters required for localization -- 3D DoDs/DoAs and TDoAs -- are now available, except the path order which determines whether to discard an estimated path or not, as only LOS and first-order reflections will be used for localization. The path classification problem is addressed in Sec. \ref{approach_path_class}.
\captionsetup[table]{}
\begin{table}[!h]\centering
\resizebox{\linewidth}{!}{
	\begin{tabular}{m{.32\linewidth}l}%{p{.25\linewidth}p{.75\linewidth}}
	\arrayrulecolor{black}\toprule
		\multicolumn{1}{c}{Method}             & \multicolumn{1}{c}{Complexity}                                                                                              \\ \midrule
		Conventional OMP [57] & $\cO\left(N_{\rm est}N_sQ\prod_{k=1}^5N_k^\rms N_k^\rma\right)$                 \\ \arrayrulecolor{lightgray}\hdashline
		MOMP [32,52]                         & $\cO\left(N_{\rm est}N_sQN_{\rm iter}(\sum_{k=1}^5N_k^\rma)(\prod_{k=1}^5 N_k^{\rms})\right)$                                                       \\ \arrayrulecolor{lightgray}\hdashline
		\textbf{Two-stage MOMP (Proposed) }             & $\cO\left(N_{\rm est}N_sQN_{\rm iter}(\sum_{k=1}^3N_k^\rma)(\prod_{k=1}^3N_k^\rms)\right)$ \\\arrayrulecolor{black}\bottomrule
	\end{tabular}
}
\caption{{Complexity comparisons for various channel estimation algorithms. }}
\label{chanEstComplex}
\end{table}
%\begin{figure}[t!]
%	\centering
%	\includegraphics[width=.8\columnwidth]{Figs/correlate.pdf}
%	\caption{Correlation between each two of the channel parameters.}
%	\label{CorrChanParam}
%\end{figure}
\begin{figure}[t!]
\centering
\includegraphics[width=.8\columnwidth]{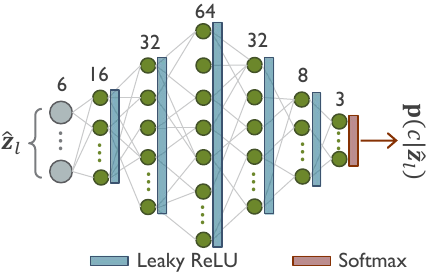}
\caption{Architecture of \textit{PathNet}.}
\label{classify_DNN_arch}
\vspace*{-5mm}
\end{figure}

\subsection{\textit{PathNet} for path classification}\label{approach_path_class}
To obtain the user position given the channel path parameters we can exploit different geometric relationships for the LOS and NLOS cases. In general, only the parameters of the LOS and first-order paths are leveraged for localization. By exploiting the laws of physics, it is possible to define a mathematical model to decide if a channel path is a LOS or a first-order reflection given their parameters (see \cite{Palacios2022Eusipco} for example). In practice, when applying the model to an estimated path, there will be numerous misclassifications due to the channel estimation error. The introduction of parameters of misclassified paths into the geometric equations that exploit the first-order or LOS nature of the path for localization will lead to a high positioning error. To overcome this limitation, we design in this section a path classification network, named \textit{PathNet}, very robust to channel estimation errors.

To build a suitable network, we study the correlations between the channel parameters and the path order, and as expected, $|\alpha|$, TDoA $\tau$, and azimuth and elevation DoAs/DoDs $\theta^{\rmx}$, $\theta^{\rmy}$, $\phi^{\rmx}$, $\phi^{\rmy}$ are related to the path order, while the phase of the path is uncorrelated with its order. Therefore, we define the input of the network to be the normalized version of all the path parameters but the phase, denoted as $\bz=[|\alpha|^2, \tau, \theta^{\rmx}, \theta^{\rmy}, \phi^{\rmx}, \phi^{\rmy}]$. For localization purposes, every estimated path $\hat{\bz}_{\ell}$ must be classified into one of the three following categories: LOS ($c=1$), first-order reflections ($c=2$), or others ($c=3$). This classification is performed given the probability vector output from \textit{PathNet}, which is defined as
\begin{equation}
\bp(c|\hat{\bz}) = \mathcal{F}(|\alpha|^2, \tau, \theta^{\rmx}, \theta^{\rmy}, \phi^{\rmx}, \phi^{\rmy}; {\boldsymbol\mu}),
\end{equation}
where $[\bp(c|\hat{\bz})]_{i}=p(c=i|\hat{\bz},i\in\{1, 2, 3\})$, $\mathcal{F}(\cdot)$ represents the operations performed by \textit{PathNet}, and ${\boldsymbol\mu}$ represents the network parameters to be trained. Hence, among $N_{\rm est} $ estimated paths of a channel, the LOS and first-order reflections can be identified according to
\begin{equation}
\hat{c}(\hat{\bz}) = \arg\max\limits_{i\in\{1,2,3\}}p(c=i|\hat{\bz}).
\end{equation}
Unlike images, the input parameters to PathNet do not exhibit visible local features, and unlike in \ac{NLP} problems, they do not contain context-sensitive or historical information. Therefore,  we simply adopt \ac{FC} layers as the major components of the network, which provide a low complexity solution to learning non-linear combinations of features embedded in the input. The proposed architecture is shown in Fig. \ref{classify_DNN_arch}. To select an effective loss function to train the network, we notice first that a higher penalization needs to be applied when classifying a second or high-order path as LOS/first-order reflection, since this would significantly deteriorate the localization performance. Regarding the misclassification of LOS/first-order reflections as high-order paths, a lower penalization should be applied, since they usually indicate an inaccurate channel parameter estimation, so that it is beneficial to discard those paths for localization. With this in mind, we propose a weighted cross-entropy loss instead of a regular cross-entropy loss to adjust the penalties, i.e.
\begin{equation}
\mathcal{L}({\boldsymbol\mu})=-e^{-\eta (c({\bf z})-\hat{c}({\bf z}))}\cdot [\bp(c|\bz)]_{c(\bz)},
\end{equation}
where $\eta$ is the customized weight coefficient.

\subsection{Geometric Localization}
In this section, we propose two different geometric localization strategies for the LOS and NLOS scenarios, both accounting for the clock offset between the TX and RX.  

\begin{figure}[ht!]
\centering
% \column
%	\subfloat[]{%
	%		\label{DoD_illustration}
	%	\includegraphics[height=0.22\textwidth]{Figs/DoD_illustration.pdf}}
\subfloat[]{%
	\label{LOSNLOS_loc_model}
	\includegraphics[height=0.22\textwidth]{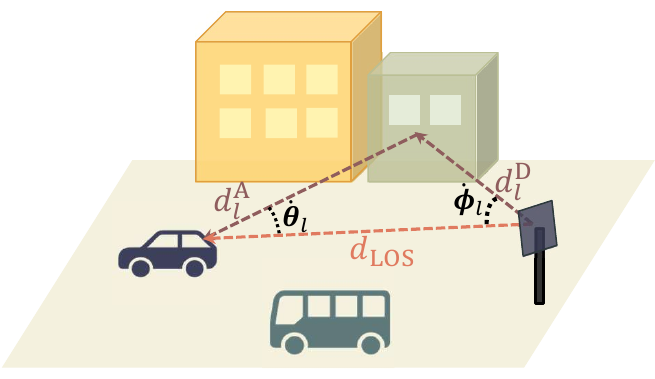}}\\
\subfloat[]{%
	\label{NLOSOnly_loc_model}
	\includegraphics[height=0.22\textwidth]{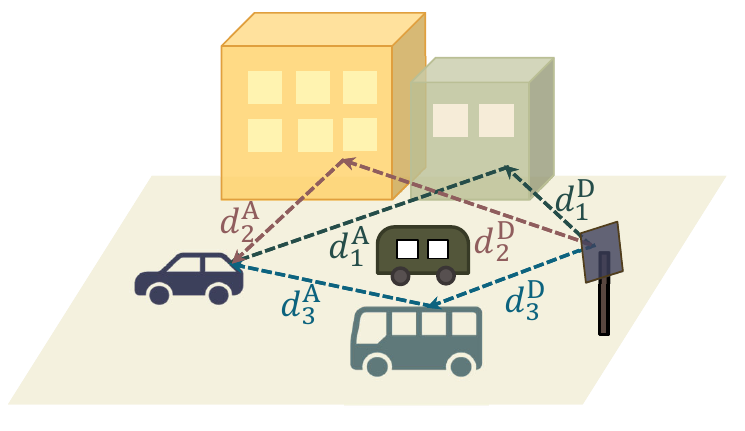}}
\caption{Illustration of the geometric relationships to be exploited for localization in (a) LOS+NLOS and (b) NLOS scenarios. }
\label{loc_model}
\end{figure}

\textbf{LOS+NLOS scenario:} We consider the geometry that can be exploited in the LOS+NLOS scenario as illustrated in Fig.~\ref{LOSNLOS_loc_model}. 
We define the angle between the DoAs of the LOS path and the $l$-th multipath component as $\dot{\theta}=\arccos({\boldsymbol \theta}_{\rm LOS}^{\sfT}{\boldsymbol \theta}_{\ell})$. Similarly, $\dot{\phi}_{\ell}=\arccos({\boldsymbol \phi}_{\rm LOS}^{\sfT}{\boldsymbol \phi}_{\ell})$ represents the angle between the DoDs. For any pair of rays composed of the LOS and any first-order reflection we can apply the Law of Sines as
\begin{equation}\label{sin_law}
\frac{d_{\rm LOS}}{\sin(\dot{\theta}_{\ell}+\dot{\phi}_{\ell})}=\frac{d_{\ell}^{\rm D}}{\sin(\dot{\theta}_{\ell})}=\frac{d_{\ell}^{\rm A}}{\sin(\dot{\phi}_{\ell})},
\end{equation}the 
where $d_{\rm LOS}=||{ \rm \bx}_{\rmt}-{\rm \bx}_{\rmr}||$ is the distance between the positions of the TX, ${ \rm \bx}_{\rmt}$, and the RX, ${\rm \bx}_{\rmr}$, which can be computed as $d_{\rm LOS}=v_{\rmc}t_{\rm LOS}$, where $v_{\rmc}$ is the speed of light and $t_{\rm LOS}$ is the time of flight; $d_{\ell}^{\rm D}$ ($d_{\ell}^{\rm A}$) is the distance between the TX (RX) and the interaction point on any surface, so that $d_{\ell}^{\rm D}+d_{\ell}^{\rm A}=v_{\rmc}t_{\ell}$. Considering these definitions we have
\begin{equation}\label{DDoF}
d_{\ell}^{\rmD}+ d_{\ell}^{\rmA}-d_{\rm LOS} = v_{\rmc}(t_{\ell}-t_{\rm LOS}) = v_{\rmc}\tau_{\ell},
\end{equation}
where $\tau_{\ell}$ is the TDoA between the $l$-th first-order reflection and the LOS. Combining \eqref{sin_law} and \eqref{DDoF}, $d_{\rm LOS}$ can be written as
\begin{equation}
\hat{d}_{\rm LOS} =\frac{v_{\rmc}\tau_{\ell} \sin(\dot{\theta}_{\ell} + \dot{\phi}_{\ell})}{\sin(\dot{\theta}_{\ell})+ \sin(\dot{\phi}_{\ell})-\sin(\dot{\theta}_{\ell} + \dot{\phi}_{\ell})}.
\end{equation}
Let's define now the vectors  ${\boldsymbol\tau}=[\tau_1,...,\tau_{L_{c=2}}]^{\sfT}$, $\dot{\boldsymbol\theta}=[\dot{\theta}_1,...,\dot{\theta}_{L_{c=2}}]^{\sfT}$, and $\dot{\boldsymbol\phi}=[\dot{\phi}_1,...,\dot{\phi}_{L_{c=2}}]^{\sfT}$.
When the number of estimated first-order reflections $L_{c=2}\geq 1$, then $\hat{d}_{\rm LOS}$ can be obtained by solving a \ac{LS}  problem with solution
{\small
\begin{equation}
	\hat{d}_{\rm LOS} =\frac{<v_{\rmc}\cdot{{\rm{\boldsymbol\tau}}} \odot\sin(\dot{\boldsymbol \theta} + \dot{\boldsymbol \phi}), \sin(\dot{\boldsymbol \theta})+ \sin(\dot{\boldsymbol \phi})-\sin(\dot{\boldsymbol \theta} + \dot{\boldsymbol \phi})>}{\|\sin(\dot{\boldsymbol \theta})+ \sin(\dot{\boldsymbol \phi})-\sin(\dot{\boldsymbol \theta} + \dot{\boldsymbol \phi})\|^2}.
\end{equation}
}
Finally, the vehicle location could be determined as
\begin{equation}
\hat{\rm\bx}_{\rmr}={\rm\bx}_{\rmt}+ \hat{d}_{\rm LOS}\cdot{\boldsymbol\phi}_{\rm LOS}.
\end{equation}
\vspace{-5mm}

\textbf{NLOS: }In this case, illustrated in Fig.~\ref{NLOSOnly_loc_model}, the geometric equations for path $l$ could be created with an extension of \eqref{DDoF} as
\begin{equation}\label{NLoS_loc_equ}
\begin{cases}
	{\bf x}_{\rmr}+{\boldsymbol \theta}_{\ell} d_{\ell}^{\rmA} = {\bx}_{\rmt} +{\boldsymbol \phi}_{\ell} d_{\ell}^{\rmD}\\
	d_{\ell}^{\rm A}+d_{\ell}^{\rmD}=\Delta d_{\ell}+d_0
\end{cases},
\end{equation}
where $\Delta d_{\ell}=v_{\rmc}(t_{\ell}-t_0)$, and $d_0=v_{\rmc}t_0$. The vehicle location can be now expressed as
\begin{equation}\label{NLOS_geo}
\bx_{\rmr}=\bx_{\rmt}+({\boldsymbol\phi}_{\ell}+{\boldsymbol\theta}_{\ell})d_{\ell}^{\rmD}-{\boldsymbol\theta}_{\ell}(\Delta d_{\ell}+d_0),
\end{equation}
with $d_{\ell}^{\rmD}$ is estimated as
\begin{equation}\label{NLOS_dD}
\hat{d}_{\ell}^{\rmD}=\frac{<{\boldsymbol\phi}_{\ell}+{\boldsymbol\theta}_{\ell}, \bx_{\rmr}- \bx_{\rmt}+{\boldsymbol\theta}_{\ell}(\Delta d_{\ell}+d_0)>}{ ||{\boldsymbol\phi}_{\ell}+{\boldsymbol\theta}_{\ell}||^2}.
\end{equation}
Now we substitute $d_{\ell}^{\rmD}$ in \eqref{NLOS_geo} with the expression in \eqref{NLOS_dD}, and define  $\Theta_{\ell}=\frac{({\boldsymbol \theta}_{\ell}+{\boldsymbol \phi}_{\ell})({\boldsymbol \theta}_{\ell}+{\boldsymbol \phi}_{\ell})^{\sfT}}{||{\boldsymbol \theta}_{\ell}+{\boldsymbol \phi}_{\ell}||^2}$. Considering these definitions, the vehicle position can be expressed now as
\begin{equation}
\bx_{\rmr}=(\bI-\Theta_{\ell})\bx_{\rmt}+\Theta_{\ell}\bx_{\rmr}-(\bI-\Theta_{\ell}){\boldsymbol\theta}_{\ell}(\Delta d_{\ell}+d_0),
\end{equation}
or alternatively
\begin{align}
(\bI-\Theta_{\ell})(\bx_{\rmr}+{\boldsymbol\theta}_{\ell} d_0)=(&\bI-\Theta_{\ell})[\bI,{\boldsymbol\theta}_{\ell}][\bx_{\rmr};d_0]\\ &=(\bI-\Theta_{\ell})(\bx_{\rmt}-{\boldsymbol\theta}_{\ell} \Delta d_{\ell}).
\end{align}
A least square estimation problem can be formulated, i.e., $[\hat{\bf x}_{\rmr};\hat{d}_0] = {\bf A}^{-1}{\bf b}$, where 
$$
\begin{cases}
{\bf A} = \sum_{l=1}^{L_{c=2}}[\bI,{\boldsymbol \theta}_{\ell}]^{\sfT}(\bI -\Theta_{\ell})[\bI,{\boldsymbol \theta}_{\ell}]\\ 
{\bf b} = \sum_{l=1}^{L_{c=2}}[\bI,{\boldsymbol \theta}_{\ell}]^{\sfT}(\bI-\Theta_{\ell})({\bf x}_{\rmt}-{\boldsymbol \theta}_{\ell}\Delta d_{\ell})
\end{cases},
$$
to obtain  the 3D vehicle position $\bx_{\rmr}$ and the clock offset. Because the rank of $\Theta_{\ell}$ is $1$ which leads to matrix $(\bI-\Theta_{\ell})$ being rank 2, and the matrix $[\bI,{\boldsymbol\theta}_{\ell}]$ is of rank $3$, the rank of $(\bI-\Theta)[\bI,{\boldsymbol\theta}_{\ell}]$ is $\min\{2,3\}=2$. Accordingly, the solution of the least square estimation problem is unique when at least $3$ estimated first-order reflections are present. In addition, after computing $\hat{\bf x}_{\rmr}$, the location of the reflection points can be determined by introducing $\hat{\bf x}_{\rmr}$ into \eqref{NLoS_loc_equ}. 

For both the LOS+NLOS and NLOS scenarios, multiple combinations of paths could exist, and will yield to different location estimates. In such a case, iterating over various combinations of paths and removing illogical localization results, e.g., those with unrealistic height estimations $\hat{x}_{\rmr}^{\perp}=[\hat{\bx}_{\rmr}]_3$, can lead to better localization results. %In the vehicular setting, we care particularly about the 2D location estimate $\hat{\bx}_{\rmr}^{\shortparallel}=[\hat{\bx}_{\rmr}]_{:2}$, which is the XY-plane location vector, for performance evaluations in the following sections.

%% sub-section: "ChanFormer" for localization refinement
\begin{figure}[!t]
\centering
\includegraphics[width=0.6\columnwidth]{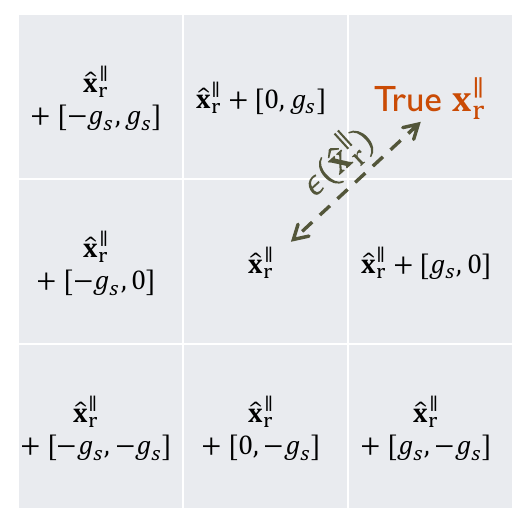}
\caption{Illustration of a $N_g\times N_g=3\times 3$ tile structure for position refinement.}
\label{tile_structure}
\vspace{-3mm}
\end{figure}
\vspace*{-4mm}
\subsection{\textit{ChanFormer} for localization refinement}
Instead of designing a network to solve the challenging regression problem of estimating the user position given the received signal, or even the channel parameters, we consider the design of a network for position refinement after obtaining an initial estimate by geometric localization. With this approach in mind, we will formulate the position refinement problem as a classification task, which is usually less challenging for an ML-based approach. To this aim, we consider a $N_g\times N_g$ tile structure with a specified grid size $g_s$ around the initial 2D location estimate $\hat{\bx}^{\shortparallel}_{\rmr}$ (obtained from the 3D location estimate $\hat{\bx}_{\rmr}$ provided by geometric localization), as shown in the example in  Fig.~\ref{tile_structure}. Our goal in this section is to create a network called \textit{ChanFormer} that obtains the probability of a tile containing the true location $\bx^{\shortparallel}_{\rmr}$. Mathematically,
\begin{equation}
p\left(\widetilde{\bx}_{\rmr}^{\shortparallel}|\hat{\bZ}\right)=p\left(\bx_{\rmr}^{\shortparallel}=\widetilde{\bx}_{\rmr}^{\shortparallel}|\hat{\bZ}, \widetilde{\bx}_{\rmr}^{\shortparallel}=\hat{\bx}^{\shortparallel}_{\rmr}+[n_{\rmx}g_s, n_{\rmy}g_s]^{\sfT}\right),
\end{equation}
where $\hat{\bZ}=[\hat{\bz}_1;...;\hat{\bz}_{N_{\rm est}}]$ is the estimated channel containing $N_{\rm est}$ estimated paths, and $|n_{\rmx}|,|n_{\rmy}|\in\left\{0,1,...,\frac{N_g-1}{2}\right\}$. $p(\widetilde{\bx}_{\rmr}^{\shortparallel}|\hat{\bZ})$ should be negatively related to the distance between $\widetilde{\bx}_{\rmr}^{\shortparallel}$ and $\bx_{\rmr}^{\shortparallel}$, which is formulated as :
\begin{equation}\label{dist_prob_model}
p\left(\widetilde{\bx}_{\rmr}^{\shortparallel}|\hat{\bZ}\right)=\frac{1}{1+e^{-\gamma\left(1-\frac{||\widetilde{\bx}_{\rmr}^{\shortparallel}-\bx_{\rmr}^{\shortparallel}||}{\delta}\right)}},
\end{equation}
where $\gamma$ is the belief factor, and $\delta$ is the scale factor for the distance $\epsilon(\widetilde{\bx}_{\rmr}^{\shortparallel})=||\widetilde{\bx}_{\rmr}^{\shortparallel}-\bx_{\rmr}^{\shortparallel}||$. 
% An illustration of the effects of $\gamma$ and $\delta$ can be found in Fig. \ref{distance_prob_model}. 
\begin{figure*}[!t]
\centering
\includegraphics[width=0.8\textwidth]{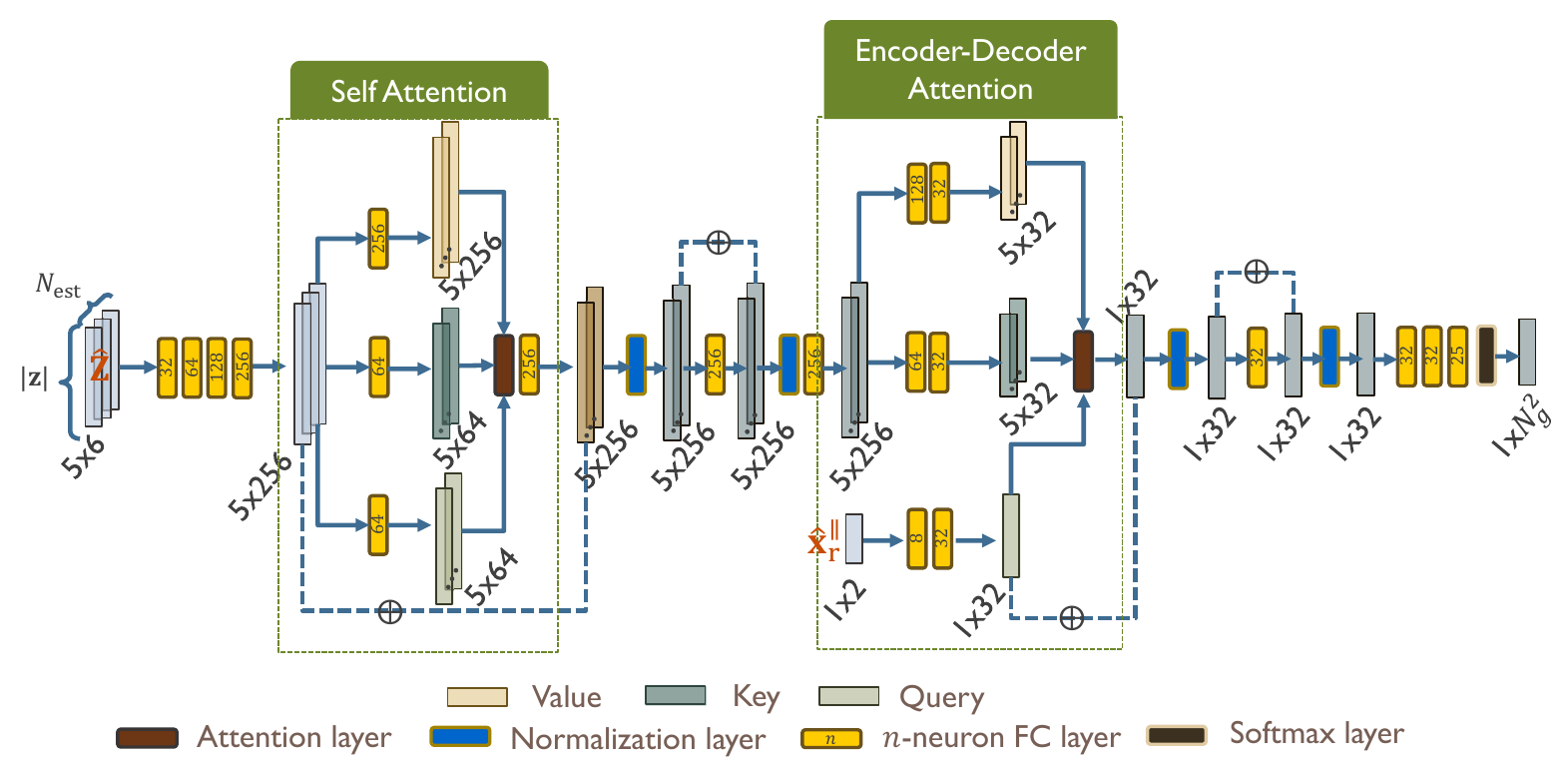}
\caption{Architecture of \textit{ChanFormer}. The self-attention block extracts the intra and crossover features of the input estimated paths. The encoder-decoder block analyzes the relationship between the initial location estimate and the extracted features of the estimated paths. }
\label{ChanFormer}
\vspace{-2mm}
\end{figure*}
\textit{ChanFormer} is meant to analyze $\hat{\bx}^{\shortparallel}_{\rmr}$ and its surroundings $\widetilde{\bx}_{\rmr}^{\shortparallel}$ to find the one that most likely meets the current estimated channel condition, i.e., $\max_{\widetilde{\bx}_{\rmr}^{\shortparallel}}p(\hat{\bZ}|\widetilde{\bx}_{\rmr}^{\shortparallel})$. The entire network can be formulated as
\begin{equation}
\hat{\bP}\left(\hat{\bZ}, \hat{\bx}_{\rmr}^{\shortparallel}\right)=\mathcal{T}\left(\hat{\bZ}, \hat{\bx}_{\rmr}^{\shortparallel}; {\boldsymbol\omega}\right)\in \mathbb{R}^{1\times N_g^2},
\end{equation} 
where $\boldsymbol\omega$ is the network parameters to be trained. Inspired by the idea of the original \textit{Transformer} \cite{Transformer2017}, the core concept of \textit{ChanFormer} is an encoder for \textit{Self-Attention} to extract features of the input estimated channel $\hat{\bZ}$, and a decoder to analyze the relationships between the estimated channel features and the initial location estimate $\hat{\bx}_{\rmr}^{\shortparallel}$ using \textit{Encoder-Decoder Attention}. The proposed architecture is shown in Fig. \ref{ChanFormer}.

\textbf{Encoder: }The workflow starts with \ac{FC} layers embedding the input estimated paths to vectors with a length of $256$. Then, the self-attention process begins by creating three abstraction matrices -- \textit{query}, \textit{key}, and \textit{value} matrices, denoted as $\bQ$, $\bK$, and $\bV$ respectively, where each row of the matrices corresponds to an estimated path. Conceptually, each $\hat{\bz}_{\ell}$ now has a high-dimensional interpretation of its features in its \textit{value} $[\bV]_{l,:}$, which can be indexed by its \textit{key} $[\bK]_{l,:}$. The following attention layer then evaluates the relationships among the paths by
\begin{equation}
{\rm Attention}(\bQ, \bK, \bV)={\rm softmax}\left(\frac{\bQ\bK^{\sfT}}{\sqrt{d_{\rmk}}}\right)\bV,
\end{equation}
where $d_{\rmk}$ is the dimension of $[\bK]_{l, :}$, and $\rm softmax$ is for atoms along $\rm axis=2$. The softmax score determines how much true channel information is represented by the $l$-th estimated path by examining the correlation between $\hat{\bz}_{\ell}$ and all the paths in $\hat{\bZ}$. It is expected that $\hat{\bz}_l$ will have the highest softmax score with itself, but other paths that have quite accurate estimations will also be assigned a relatively high score. Therefore, the $\rm Attention$ output is the expression of each path that integrates information from all other paths. The less reliable estimated paths will have a smaller influence on the expressions. This means that the attention mechanism emphasizes more accurate paths in this step, so that the true channel is better represented, regardless of the presence of the noises from the channel estimation process and/or the misclassified paths. In addition, the attention layer is capable of analyzing the input without being constrained by its chronicle orders, which exceeds the capabilities of the convolutional layer that considers path relationships within a fixed window, or the \ac{FC} layer that relies on connections of all the input parameters.

\textbf{Decoder: }The input $\hat{\bx}_{\rmr}^{\shortparallel}$ serves as a \textit{query}, referring to which the network generates the probability map $\bP(\hat{\bZ})^{\star}\in \mathbb{R}^{N_g\times N_g}$ of the tiles with $\hat{\bx}_{\rmr}^{\shortparallel}$ at the center. Note that, though $\hat{\bx}_{\rmr}^{\shortparallel}$ is the only input at the decoder, the network actually evaluates all candidate locations within the tiles given the grid size. In this part, the attention layer improves the initial location estimate accuracy by assigning a higher probability to a candidate location $\widetilde{\bx}_{\rmr}^{\shortparallel}$ that aligns better with the channel representation obtained from the encoder. Note that the output $\hat{\bP}(\hat{\bZ}, \hat{\bx}_{\rmr}^{\shortparallel})$ is reshaped to $\bP(\hat{\bZ}, \hat{\bx}_{\rmr}^{\shortparallel})^{\star}\in\mathbb{R}^{N_g\times N_g}$ to acquire the probability map to simplify the process of accessing $\widetilde{\bx}_{\rmr}^{\shortparallel}$ associated with $p(\widetilde{\bx}_{\rmr}^{\shortparallel}|\hat{\bZ})$. By referring to the tile with the highest probability:
\begin{equation}
\begin{aligned}
	[j^{\star}, i^{\star}] &= \arg\max\limits_{j, i}\left[\hat{\bP}(\hat{\bZ}, \hat{\bx}_{\rmr}^{\shortparallel})^{\star}\right]_{j,i}
	\Rightarrow\\ &[n_{\rmx}^{\star}, n_{\rmy}^{\star}]=\left[i^{\star}-\frac{N_g+1}{2}, \frac{N_g+1}{2}-j^{\star}\right],
\end{aligned}
\end{equation}
the refined location is given by:
\begin{equation}
\widetilde{\bx}_{\rmr}^{\shortparallel(\star)}=\hat{\bx}^{\shortparallel}+[n_{\rmx}^{\star}g_s, n_{\rmy}^{\star}g_s]^{\sfT}.
\end{equation}

To train the network, we evaluate both MSE loss and Kullback-Leibler (KL) divergence loss for learning the probability map distribution.

\vspace*{-3mm}
\section{Simulation Results}\label{sec:SimResults}
In this section, we present simulation results to evaluate the performance of the different modules designed in this paper. We begin by detailing the simulation setup in Sec. \ref{Simulation setups}. Then, in Sec. \ref{chanest_results}, we assess the accuracy of the MOMP-based channel estimation stage. In Sec. \ref{PathNet_result}, we demonstrate the results of path classification using \textit{PathNet}. We finally discuss the localization results, including the initial estimation only exploiting geometric relationships and the enhanced performance after applying \textit{ChanFormer}. 
{Note that, \textit{PathNet} is exclusively trained on perfect channels, while the two-stage MOMP based channel estimates are used in the testing stage for \textit{PathNet}, and both training and testing for \textit{ChanFormer}. As channel estimation errors  propagate through the system, we analyze their impact on the different stages of our solution, including path order classification, geometric localization, and position refinement using \textit{ChanFormer}.}

%\begin{figure}[t!]
%	\centering
%	\includegraphics[width=.9\columnwidth]{Figs/Sim_Env.pdf}
%	\caption{Ray-tracing simulation of an urban environment with cars and trucks distributed across $4$ lanes. }
%	\label{Sim_Env}
%\end{figure}
\vspace*{-2mm}
\subsection{Simulation setup}\label{Simulation setups}
\textbf{Ray-tracing simulation for realistic channels: }
We run $2500$ electromagnetic simulations of a vehicular environment in Rosslyn City, Virginia, on a $240\times120$ m$^2$ plane, using \textit{Wireless Insite} software\cite{WirelessInsiteSW}. %, as the snapshot shown in Fig. \ref{Sim_Env}. 
In each simulation, around $30$ vehicles are randomly distributed across the four lanes for initial access, with $80\%$ being cars and $20\%$ being trucks according to the 3GPP methodology for simulation of vehicular communication systems \cite{3GPPVehEnv}. The \ac{BS} is located at $\bx_{\rmt}=[120, -21, 5]$ m, down facing the road. Parameters for materials of the building/territorial surfaces, the vehicle sizes, placements of antennas on the vehicle and BS, etc., follow the deployments in \cite{anum2020passive}. $4$ active cars are randomly selected to communicate with the BS in the $73$ GHz band. With each car equipped with $4$ communication arrays, the simulations provide $4\times 4\times 2500=40$k channels as the dataset $\mathcal{S}$, and every channel has a maximum of $L=25$ multipath components. The first $24$k channels denoted as $\cS_{\rm tr}$ are split into $3:1$ for training and validations, and the remaining $16$k channels serve as the testing set $\mathcal{S}_{\rm te}$ for all the performance evaluations. 

\textbf{Communication system:}  In this paper, we use an antenna setting of $N_{\rmt} = N_{\rmt}^{\rmx}\times N_{\rmt}^{\rmy} = 16\times 16$, $N_{\rmr} = N_{\rmr}^{\rmx}\times N_{\rmr}^{\rmy} = 8\times 8$, and a transmitted power of $P_{\rmt}=40$ dBm. The number of RF chains at TX and RX are set to be $N_{\rmt}^{\rm RF}=8$ and $N_{\rmr}^{\rm RF}=4$. The communication system operates at a carrier frequency $f_c=73$ GHz with a bandwidth $B_c=1$ GHz. 
A noise power $\sigma^2_{\bn}=-84$ dBm is computed using $T=288^{\degree}$F. Given the \ac{RMS} delay-spread of the simulated channels and the bandwidth, the number of delay taps is fixed to $N_{\rmd}=64$. $N_s=\min\{N_{\rmt}^{\rm RF}, N_{\rmr}^{\rm RF}\}=4$ training data 
streams with a length of $Q=64$ are transmitted. We use the raised-cosine filter with a roll-off factor of $0.4$ to simulate pulse shaping and other filtering effects in the discrete equivalent channel.

\vspace*{-3mm}
\subsection{MOMP based low complexity 3D channel estimation}\label{chanest_results}
% settings for MOMP channel estimation
The training matrix $\bF$ is constructed by the Khatri-Rao product of the precoders along the azimuth and elevation planes, i.e., $\bF = \bF^{\rmx}\circ \bF^{\rmy}$. Each column of $\bF^{\rmx}$ ($\bF^{\rmy}$) is extracted from the DFT codebook of size $N_{\rmt}^{\rmx}$ ($N_{\rmt}^{\rmy}$), e.g., $\forall [\bF^{\rmx}]_{:,i}\in \left\{\ba'(\varphi)|\varphi=0,\frac{2\pi\cdot 1}{N_{\rmt}^{\rmx}},...,\frac{2\pi\cdot (N_{\rmt}^{\rmx}-1)}{N_{\rmt}^{\rmx}}\right\}$, where $\ba'(\varphi)=\frac{1}{\sqrt{N_{\rmt}^{\rmx}}}\left[0,e^{j\cdot 1\cdot \varphi},...,e^{j\cdot (N_{\rmt}^{\rmx}-1)\cdot \varphi}\right]^{\sfT}$. The same procedure applies to the combiners $\bW$. We use a setting of $M_{\rmt}=16,\ M_{\rmr}=64$, and form $\bY_M$ by collecting a total of $M=M_{\rmr}\times M_{\rmt}=1024$ frames. The size --or the resolution-- of the dictionaries is based on the number of their atoms along the dimension, with a specific constant $K_{\rm res}$ determining the proportion, i.e., $N_k^{\rma}=K_{\rm res}\cdot N_k^{\rms}$. In our case, $N_1^{\rms}=N_{\rmt}^{\rmx}$, $N_2^{\rms}=N_{\rmt}^{\rmy}$, and $N_3^{\rms}=N_{\rmd}$. The impact of $K_{\rm res}$ is studied in \cite{MOMP2022}. Here we set $K_{\rm res}=128$ as it brings a comparable performance to using a higher resolution setting, such as $K_{\rm res}=1024$, while being computationally more efficient. 

\begin{figure}[ht!]
\centering
\subfloat[]{%
	\label{chan_est_angle}
	\includegraphics[width=0.8\linewidth]{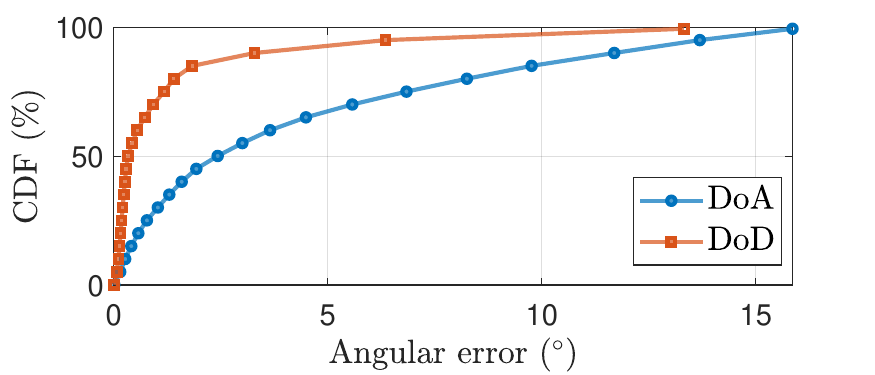}}\vspace{-.6mm}
\subfloat[]{%
	\label{chan_est_delay}
	\includegraphics[width=0.8\linewidth]{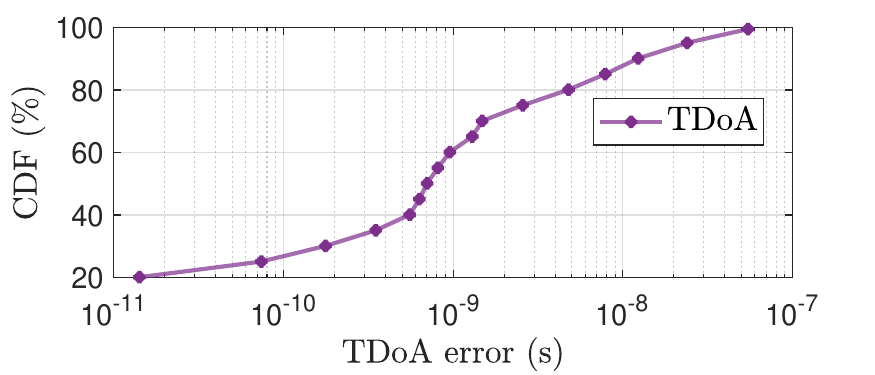}}
\caption{MOMP-based channel estimation performance using a setting with a $16\times 16$ array at the TX and a $8\times 8$ array at the RX. Plots acquired based on the whole dataset $\cS$.}
\label{chan_est_err}
\end{figure}
% \begin{figure}[t!]
% 	\centering
% %	% \column
% %	\subfloat[]{%
	% 		\label{chan_est_angle_err}
	% 		\includegraphics[width=0.9\columnwidth]{Figs/chan_est_errs.pdf}
	% %	}\\
% %	\subfloat[]{%
	% %		\label{chan_est_delay_err}
	% %		\includegraphics[width=0.9\columnwidth]{Figs/chan_est_delay_err.pdf}}
% 	\caption{\yuncommentY{(Replot)} MOMP-based channel estimation performance using a setting with a $16\times 16$ array at the TX and a $8\times 8$ array at the RX. Plots acquired based on the whole dataset $\cS$.}
% 	\label{chan_est_err}
% 	\vspace{-2mm}
% \end{figure}

The angle and delay estimation performance is in Fig.~\ref{chan_est_err}. The estimation errors are calculated by matching an estimated path to its closest true path in the channel. We obtain  DoD estimaties with an average error of $0.5^\degree$, and DoA estimates with an average error of $2.5^{\degree}$. This is reasonable as the TX is equipped with a $16\times 16$ antenna array, while the RX array size is $8\times 8$. In addition, to reduce complexity, the DoA is extracted after DoD estimation, which further reduces the ability of the algorithm to provide high accuracy. An average delay error of $1e-9$ s is also observed. Note that these results include all channel paths, not only first-order reflections, and are deteriorated by the larger estimation error on second or high-order paths. However, this does not reflect the localization accuracy, as \textit{PathNet} will identify the LOS and first-order reflections for localization. We do not introduce any comparison to prior work since there is no existing algorithm that considers the filtering effects, performs time domain estimation so the delays can also be extracted,  and can run with the realistic array sizes used in our setup. This is because the memory and computational complexity requirements of the other approaches exceed what a current personal computer or server can provide. 

%\begin{figure}[t!]
%	\centering
%	\includegraphics[width=\columnwidth]{Figs/database_stat.pdf}
%	\caption{Distribution of the power gap $\Delta|\alpha|^2$ between the LOS and strongest 1st order path of $14$k valid LOS channels from the entire dataset containing $40$k channels. $40\%$ channels are with $\Delta|\alpha|^2>30$ dB.}
%	\label{database_stat}
%\end{figure}

\subsection{\textit{PathNet} for LOS and first-order reflection identifications}\label{PathNet_result}
\begin{figure}[ht!]
\centering
\subfloat[]{%
	\label{cnfsMap_real}
	\includegraphics[height=0.45\linewidth]{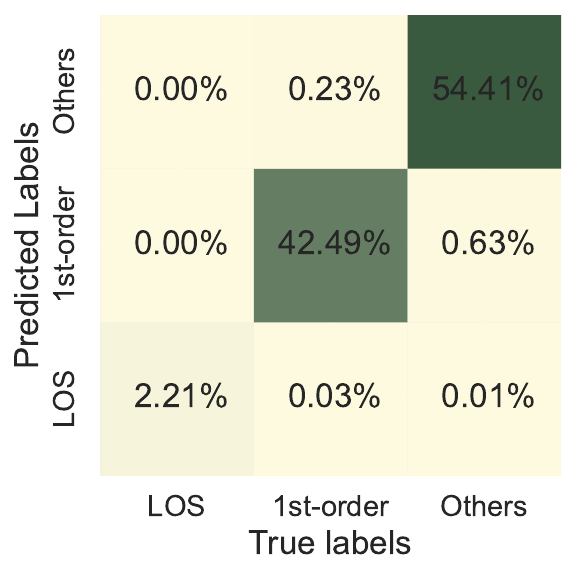}}\hfill
\subfloat[]{%
	\label{cnfsMap_MOMPest}
	\includegraphics[height=0.45\linewidth]{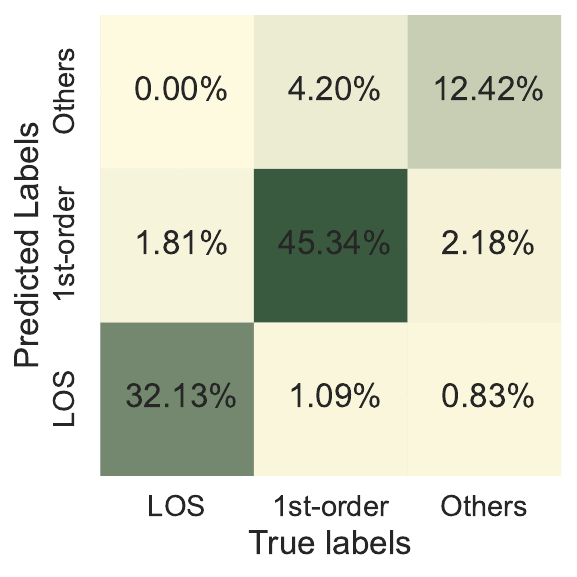}}
\caption{{Path order classification performance with \textit{PathNet}. (a) Classification with perfect channel parameters (20 paths per channel); (b) Classification with MOMP estimated channels (5 estimated paths per channel).}}
\label{cnfsMap}
\end{figure}

\textit{PathNet} is trained based on $\cS_{\rm tr}$, which lasts for $1000$ epochs with an early stopping depending on the convergence of the validation loss. The customized weight for tweaking the penalty in $\mathcal{L}({\boldsymbol\mu})$ is set to $\eta=0.2$. We adopt Adam optimizer \cite{kingma2014adam}, and set the learning rate $1e-3$ with a decay rate of $0.95$ every $200$ training epochs. {The path classification performance represented by confusion matrices in Fig.~\ref{cnfsMap} is evaluated with channels in $\mathcal{S}_{\rm te}$, where Fig.~\ref{cnfsMap_real} and Fig.~\ref{cnfsMap_MOMPest} show the path order classification results for true and MOMP estimated channels, respectively. With the perfect channel parameters, the classification accuracy reaches $\sim 99\%$, highlighting the generalization capability of the simple yet effective network. When using MOMP estimated channels, the classification accuracy reduces to $94.7\%$ for LOS, $90.0\%$ for first order reflections, and $80.5\%$ for other paths.  That being said, paths are more likely to be misclassified as higher order paths rather than first-order reflections or LOS, which are subsequently discarded for localization. In the rare instances where high order paths are misclassified as LOS or first-order reflections, we have incorporated mitigation strategies. These include disregarding paths that yield anomalous height estimates, and using the LOS with the highest power when multiple LOS paths are identified.}

% \begin{table}[h!]
% 	\centering 
% 	\resizebox{.9\columnwidth}{!}{
	% 		\begin{tabular}{lp{0.24\columnwidth}p{0.24\columnwidth}p{0.24\columnwidth}}
		% 			\toprule
		% 			\textbf{Overall}   & \textbf{LOS} ($c=1$)      & \textbf{1st order path} ($c=2$) & \textbf{Other} ($c=3$)    \\
		% 			\midrule
		% 			$98.81\%$ & $99.86\%$ & $99.13\%$             & $98.67\%$ \\
		% 			\bottomrule
		% 		\end{tabular}
	% 	}
% 	\caption{Categorical path classification accuracy based on $16$k testing channels ($25$ paths simulated for each channel). (Add channel estimation impact on PathNet)}
% 	\label{path_class_perform}
% 	\vspace{-3mm}
% \end{table}
\captionsetup{}
\begin{figure}[t!]
\centering
\includegraphics[width=0.9\columnwidth]{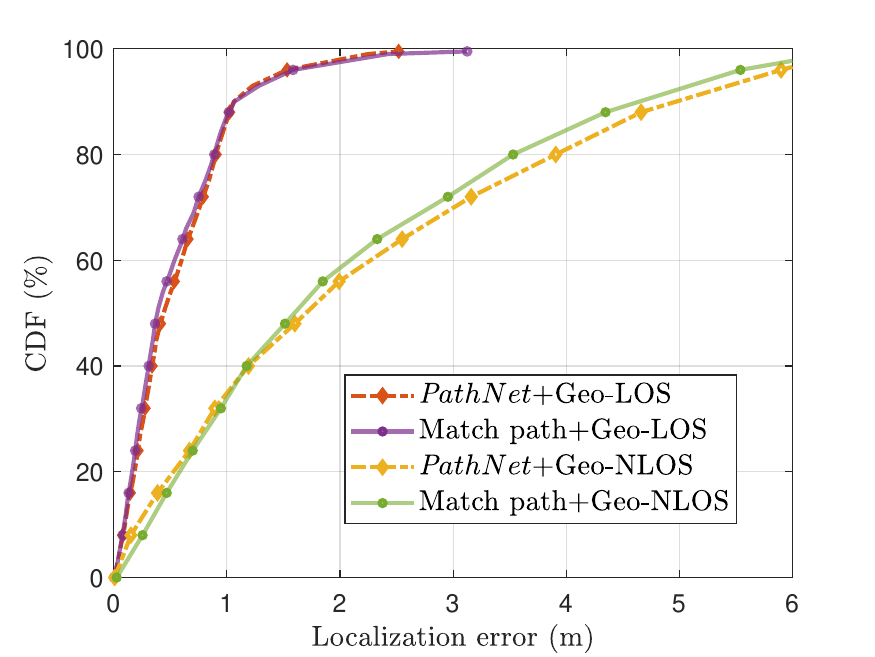}
\caption{{Geometric localization performance for MOMP estimated channels in $\cS^+_{\rm te}$. ``Geo-LOS" and ``Geo-NLOS" refer to LOS+NLOS and NLOS-only localization methods. Results with path orders determined by matching the estimated paths to their closest counterparts in the true channel (denoted as ``Match path") are included for comparison.}
}
\label{ini_perform}
\vspace{-2mm}
\end{figure}
% \begin{table}[!b]
% 	\centering
% 	\resizebox{\columnwidth}{!}{
	% 		\begin{tabular}{lllll}
		% 			\toprule
		% 			% colors: FFF0C8
		% 			\textbf{Channel}                  & \textbf{$5$th}  & \textbf{$50$th} & \textbf{$80$th} & \textbf{$95$th} \\ \midrule
		% 			Overall            & $0.06$ & $0.51$ & $0.92$ & $1.70$ \\ 
		% 			\rowcolor[HTML]{FFF6CC} 
		% 			Overall$^{\star}$  & $0.04$ & $0.22$ {  \scriptsize $57\%\downarrow$} & $0.32$ {  \scriptsize $65\%\downarrow$} &  $0.72$ {  \scriptsize $58\%\downarrow$} \\ \hline
		% 			LOS+NLOS           & $0.06$ & $0.49$ & $0.88$ & $1.26$ \\
		% 			\rowcolor[HTML]{FFF6CC} 
		% 			LOS+NLOS$^{\star}$ & $0.04$ & $0.21$ {  \scriptsize $57\%\downarrow$} & $0.31$ {  \scriptsize $65\%\downarrow$} & $0.45$ {  \scriptsize $64\%\downarrow$} \\\hline
		% 			NLOS               & $0.21$ & $1.67$ & $3.87$ & $5.97$ \\
		% 			\rowcolor[HTML]{FFF6CC} 
		% 			NLOS$^{\star}$     & $0.09$ & $0.75$ {  \scriptsize $55\%\downarrow$} & $3.63$ & $5.80$ \\ 
		% 			\bottomrule
		% 		\end{tabular}
	% 	}
% 	\caption{\yuncommentY{(Pending update)} Percentile statistics of 2D localization errors (m) with and without \textit{ChanFormer}. $\star$: location estimates refined by \textit{ChanFormer}, which show a significant error reduction; ``Overall" shows the merged localization results of using channels in $\mathcal{S}^+_{\rm te}$. }
% 	\label{all_perform}
% \end{table}
\begin{table}[!b]
\centering
\resizebox{\columnwidth}{!}{
	\begin{tabular}{p{.4\linewidth}p{.1\linewidth}p{.1\linewidth}p{.1\linewidth}p{.1\linewidth}m{.2\linewidth}<{\centering}}
		\toprule
		Method                      & $5$th           & $50$th          & $80$th          & $95$th          & $p(\epsilon<1\text{ m})$ \\ \midrule
		% \rowcolor[HTML]{FFF6CC} 
		\textit{PathNet}$+$Geo-LOS             & $0.05$          & $0.44$          & $0.90$          & $1.42$          & $87\%$                                   \\ \arrayrulecolor{lightgray}\hdashline
		\textit{PathNet}$+$Geo-NLOS            & $0.10$          & $1.73$          & $3.90$          & $5.73$          & $36\%$                                   \\ \arrayrulecolor{black}\midrule
		% \rowcolor[HTML]{FFF6CC}  
		\textit{PathNet}$+$Geo-LOS $+$\textit{ChanFormer}  & $\bf 0.04$  & $\bf 0.18$ {  {\scriptsize ($59\%\downarrow$)}} & $\bf 0.28$ {  {\scriptsize ($69\%\downarrow$)}} & {$\bf 0.58$} {  {\scriptsize ($59\%\downarrow$)}} & $\bf 98\%$                          \\ \arrayrulecolor{lightgray}\hdashline
		\textit{PathNet}$+$Geo-NLOS $+$\textit{ChanFormer} & $\bf 0.07$  & $\bf 0.77$ {  {\scriptsize ($55\%\downarrow$)}} & $\bf 3.09$ {  {\scriptsize ($21\%\downarrow$)}} & $\bf 5.60$ & $\bf 55\%$                          \\ \arrayrulecolor{black}\bottomrule
	\end{tabular}
}
\caption{Localization error percentiles (m) before and after applying \textit{ChanFormer}. Red percentages in brackets indicate error reduction with \textit{ChanFormer}.}
\label{all_perform}
\end{table}
\subsection{Localization performance}\label{ini_loc_results}
\subsubsection{Dataset preparation}
{Considering practical localization applications, the vehicle location is calculated based on the array with the strongest received power. Given $4$ arrays deployed per vehicle, it reduces the number of channels by $4\times$, resulting in a total of $10$k channels. We further examine the database for valid LOS channels ($L_{c=2}\geq 1$) and valid NLOS channels ($L_{c=2}\geq 3$), and establish criteria to exclude channels where the vehicle cannot be located exploiting measurements from a single BS.  For LOS channels, we measure the power gap $\Delta |\alpha|^2$ between the LOS and the strongest first-order path. We find that $40\%$ of the channels obtained with Wireless Insite include very weak first-order reflections, with $\Delta |\alpha|^2>30$ dB. We assume that if $\Delta |\alpha|^2>30$ dB for all the first-order paths, the channel is purely LOS, and the vehicle cannot be located with a single BS due to the lack of strong first-order reflections.  Analogously, for valid NLOS channels, we check the received power levels to determine a threshold to ensure a sufficient number of first-order paths ($\geq 3$) so the vehicle can be located. In particular, we require paths received with an attenuation $<40$ dB, and exclude from the database any NLOS channels containing less than $3$ qualifying paths. Therefore, the new sets $\cS_{\rm tr}^+\in\cS_{\rm tr}$ containing $4085$ LOS and $1085$ NLOS channels, and $\cS_{\rm te}^+\in \cS_{\rm te}$ containing $1385$ LOS and $375$ NLOS channels are formed. The following evaluations of the localization performance are based on $\cS_{\rm te}^+$.}

%We exclude these channels from the dataset used to evaluate our approach.
%% It is important to note that while these simulated realistic channels provide a useful representation of real-world channels, they do have limitations due to the absence of factors such as traffic lights and building windows that can produce additional reflections, some of which can be quite strong. These factors could have an impact on the overall channel characteristics. 
%In summary, we consider that a vehicle is locatable when the channel status satisfies the following rules: 1) The channel is valid with sufficient paths for localization; 2) $\Delta |\alpha|^2 \leq 30$ dB for LOS channels; 3) For each NLOS channel, the set of 1st order reflections with a received power of $P_{\rmr}\geq -80$ dBm -- $\mathcal{Z}=\left\{\hat{\bz}_l\ |\ P_{\rmr}(l)\geq -80\ {\text{dBm}}, c(\hat{\bz}_l)=2\right\}$ -- has a length of $|\mathcal{Z}|\geq 3$. It is worth noticing that the statistic is based on ideal channels rather than the estimated, hence there are likely fewer estimated channels that qualify for usage in our localization strategy.

\subsubsection{Geometric (initial)  localization performance}
Fig. \ref{ini_perform} shows the \ac{CDF} of localization error (m), and the 5, 50, 80, and 95th-percentile accuracies are presented in Table \ref{all_perform}. We observe that sub-meter accuracy localization is realized for $87\%$ of the users in LOS channels and $36\%$ of the users in NLOS channels. The compromised performance for NLOS channels is due to the small pool of qualified estimated paths and the decreased accuracy of MOMP channel estimations. Nevertheless, the achieved performance should be considered the worst-case scenario, as the real-world channels are likely to include more usable reflections with higher power from traffic lights, building windows, and other details of the vehicular scenario which are not present in our electromagnetic simulation of the environment. {To study the combined impact of channel estimation errors and \textit{PathNet} classification errors, we also include in Fig. 8 the localization results where the path orders are determined by matching the estimated paths to their closest counterparts in the true channel (denoted as ``Match path") instead of using \textit{PathNet} predictions. The performance is comparable for LOS channels, due to the very high path classification accuracy in this case.  For NLOS cases, we do observe the impact of the misclassifications and channel estimation errors in performance, since NLOS paths are weaker, their estimation accuracy is lower, and this also increases the likelihood of being misclassified.}
%

%\begin{figure}[t!]
%	\centering
%	\includegraphics[width=\columnwidth]{Figs/perform_various_tiles.pdf}
%	\caption{Overall 2D localization error (m) when training \textit{ChanFormer} on $\mathcal{S}^{+}_{\rm tr}$ using various tile settings.}
%	\label{perform_various_tiles}
%\end{figure}
\begin{figure}[ht!]
\centering
\subfloat[]{%
	\label{locAcc_vs_modelSize}
	\includegraphics[width=0.7\linewidth]{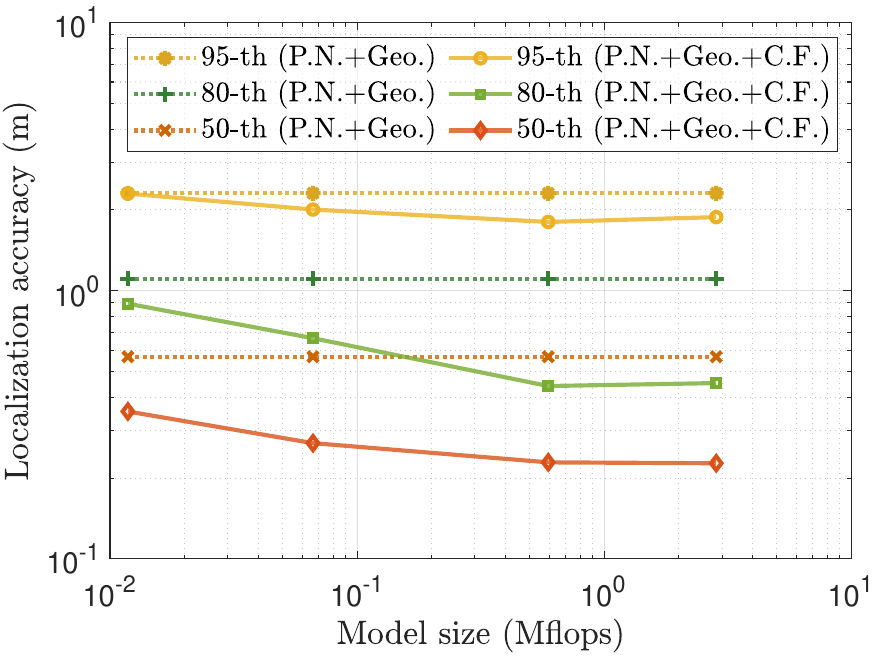}}\hfill
\subfloat[]{%
	\label{locAcc_vs_dataSize}
	\includegraphics[width=0.7\linewidth]{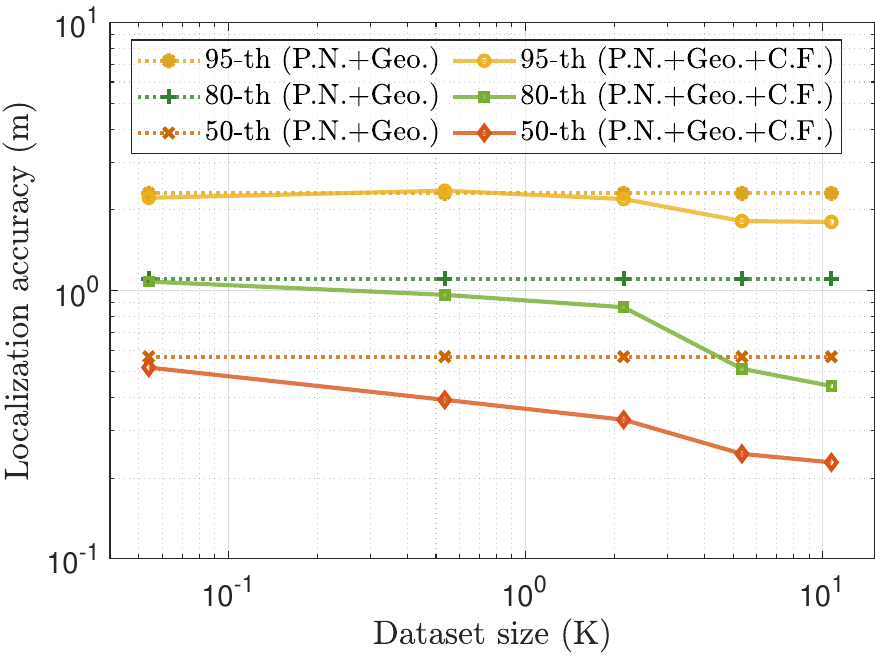}}
\caption{{Localization refinement using \textit{ChanFormer} with various model and dataset size configurations. ``P.N.+Geo." represents \textit{PathNet} and geometric localization, and ``C.F." means refinement with \textit{ChanFormer}. (a) Performance bottlenecked by model size, where a large dataset is used to ensure the model capacity is the main bottleneck; (b) Performance bottlenecked by dataset size, where the model with an optimal size is used. }}
\label{locAcc_vs_scale}
\end{figure}
\begin{figure}[t!]
\centering
\includegraphics[width=0.9\columnwidth]{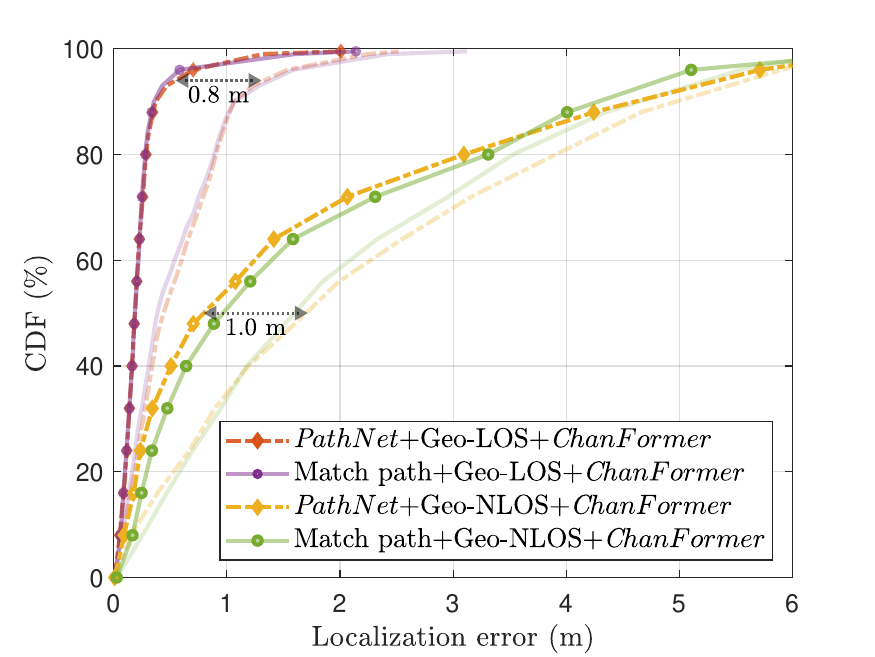}
\caption{Position estimates refined with \textit{ChanFormer}. Lines with transparency represent the results for geometric localization (Fig.~\ref{ini_perform}) as a reference. The 95th-percentile 2D error is reduced by $0.8$ m ($59\%\downarrow$) for $95\%$ of users in LOS channels, and $1$ m ($55\%\downarrow$) for half of the users in NLOS channels, realizing the expected sub-meter localization.
}
\vspace{-2mm}
\label{ChanFormer_refined}
\end{figure}

\subsection{Localization refinement with \textit{ChanFormer}}\label{refined_loc_results}
%The impacts of $\gamma$ and $\delta$ are studied as shown in Fig. \ref{distance_prob_model}, accordingly to which $\gamma=5$ and $\delta=1$ are set in Equ. (\ref{dist_prob_model}) to have a smooth and balanced probability function. A $5\times 5$ tile structure with a grid size $g_s=0.4 \text{ m}$ is adopted, which is found to be the best option among the settings shown in Fig. \ref{perform_various_tiles}. \textit{ChanFormer} is trained based on $3568$ qualified channels ($3316$ LOS / $252$ NLOS) in $\mathcal{S}^{+}_{\rm tr}$ and $\mathcal{S}^{+}_{\rm val}$, with a batch size of $32$. Adam optimizer is used, and  the learning rate is $2e-4$ with a decay rate of $0.95$ per $200$ epochs. Loss on the validation set is monitored for early stopping.

{We employ a $5\times 5$ tile structure with a grid size $g_s=0.4 \text{ m}$ as the output for \textit{ChanFormer}, which is found to be the best option among the test settings. The labels for the $5\times 5$ grids are calculated by setting $\gamma=5$ and $\delta=1$ in \eqref{dist_prob_model}, where $p(\bar{\bx}^{\sprl}_{\bR})|\bZ)$ drops to $\leq 0.6\%$ for the ranging error $\epsilon(\bar{\bx}^{\sprl}_{\bR})\geq 2$ m. This network is trained with a batch size of $64$ using the Adam optimizer with the learning rate of $2e-4$. To determine the optimal model architecture and dataset sizes, we train the network with various model and dataset configurations. The complexity of the model is varied by adjusting the number of layers and neurons per layer, allowing us to investigate the impact of model size on performance. Additionally, we assess the impact of the dataset size by adjusting the training length.  Fig.~\ref{locAcc_vs_scale} presents the localization accuracy percentiles on $\mathcal{S}_{\text{te}}^+$ for various configurations, including initial estimates to highlight \textit{ChanFormer}'s capability to improve position estimation accuracy. The resulting accuracy initially improves as the model size increases. However, a saturation point can also be observed, indicating that the model has become overly complex. On the other hand, expanding the dataset enhances the accuracy, as it provides the network with more comprehensive features to learn
}

Fig. \ref{locAcc_vs_scale} illustrates the percentile distribution of localization accuracy on $\mathcal{S}_{\text{te}}^+$ across various configurations, including initial estimates to highlight ChanFormer's capability for accuracy improvement.
% Based on different performance studies we choose $\gamma=5$ and $\delta=1$ in \eqref{dist_prob_model}, which provide a smooth and balanced location probability function. A  $5\times 5$ tile structure with a grid size $g_s=0.4 \text{ m}$ is adopted, which is found to be the best option among the test settings. This configuration of \textit{ChanFormer} is trained with a batch size of $64$ using the Adam optimizer with the learning rate of $2e-4$. 
With the optimal model and dataset size, we assess the localization refinement performance based on set $\mathcal{S^+_{\rm te}}$, and the CDF of localization errors is in Fig. \ref{ChanFormer_refined}. The performance per percentiles can be found in Table \ref{all_perform}, highlighting the accuracy improvement when applying \textit{ChanFormer} to geometry based localization results. The refinement reduces the 95th-percentile error to $0.58$ m from $1.42$ m, i.e., $59\%$ accuracy improvement, for users in the LOS. An error reduction to $0.77$ m from $1.73$ m, i.e., $55\%$ accuracy improvement, is achieved for half of the users in the NLOS scenario. In conclusion, $98\%$ of the users in LOS and $55\%$ of the users in the NLOS case achieve sub-meter accuracy. 
{The marginally better performance with \textit{PathNet} predicted orders over using matched paths can be attributed to grid resolution constraints. Localization with matching paths has slightly lower errors and undergoes less refinement compared to using \textit{PathNet} determined paths, resulting in a smaller accuracy improvement.}

%Fig. \ref{prob_maps} shows an example of pairs of ground truth and predicted probability maps in an NLOS situation that illustrates the operation of Chanformer. A reduction of 2D error from $\epsilon(\hat{\bx}_{\rmr}^{\shortparallel})=0.5 \text{ m}$ to $\epsilon(\widetilde{\bx}_{\rmr}^{\shortparallel})=0.07$ m for the LOS case, and an error reduction from $0.56$ m to $0.08$ m for the NLOS case, are obtained for that particular channel. Note that, \textit{ChanFormer} working like a filter can be effective when the initial location estimate error $\epsilon(\hat{\bx}_{\rmr}^{\shortparallel})< 2.5$ m, allowing the network to remove the noise embedded in both $\hat{\bZ}$ and $\hat{\bx}_{\rmr}^{\shortparallel}$, and pull the location estimate in the right direction, whereas it cannot effectively correct $\epsilon(\hat{\bx}_{\rmr}^{\shortparallel})$ when $\hat{\bx}_{\rmr}^{\shortparallel}$ is hopped too far away from the tile panel, especially in the NLOS case. Fig. \ref{bad_refine_case} illustrates the situation where \textit{ChanFormer} enables an error reduction of $1.13$ m, but the error remains at $1.3$ m as the initial estimate is $2.5$ m away from the true location. 

\begin{figure}[t!]
\centering
\includegraphics[width=.9\columnwidth]{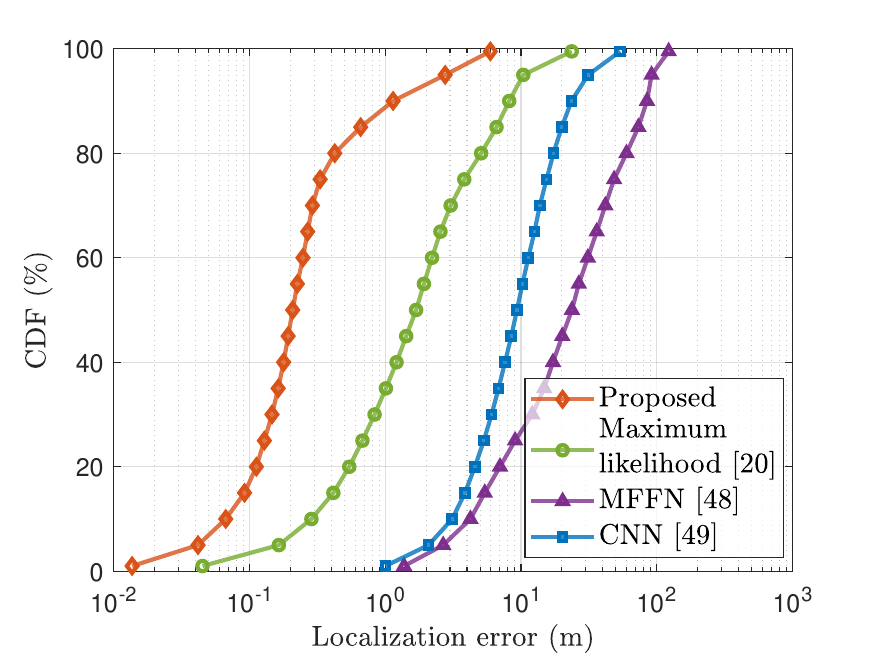}
\caption{ {Comparison of state-of-the art and proposed scheme. CDF obtained using $\mathcal{S}^{+}_{\rm te}$.}}
\label{CDF_comp_baseline}
\end{figure}
\textbf{Comparison to prior work:} Most of the previous studies mentioned in Sec. \ref{PriorWork} assume unrealistic channels and simplistic communication system settings (for example operating with the true channel parameters instead of the estimated ones, or neglecting the clock offset), limiting their performance when evaluated with our data set and system model. {To perform comparisons to prior work, we use our realistic ray-tracing simulated channels and signal model that accounts for filtering effects. We implemented three different localization strategies in prior work as baselines, one exploiting geometric localization \cite{nazari2023mmwave} and two other exploiting deep learning architectures \cite{lv2022deep, gante2020deep}. To guarantee a fair comparison, all the approaches exploit the channel parameters estimated with two-stage MOMP as described in Section III.A.3. The localization results with this experimental setting can be found in Fig. \ref{CDF_comp_baseline}.  Our hybrid model/data driven localization method significantly outperforms all the solutions in prior work, achieving sub-meter accuracy for $90\%$ of the users and errors below 30 cm for $50\%$ of the users. Note that the performance obtained with \cite{nazari2023mmwave} degrades compared to that shown in the original paper due to the use of realistic channels in our simulations instead of ideal ones -- with only LOS and first-order reflections-- and the introduction of filtering effects. Similarly, the performance degradation of the approach in \cite{lv2022deep} comes from exploiting the true (not estimated) channel parameters in the original work.}

\section{Conclusion}\label{conclu}
We developed a hybrid data/model-driven approach to obtain 3D localization in a vehicular network operating at mmWave. We considered a realistic channel model accounting for filtering effects and an unknown TX-RX clock offset. We generated realistic channel datasets for evaluation using ray-tracing. We designed \textit{PathNet}, a data driven path classification strategy to select the LOS and first-order paths from the estimated channel, achieving a classification accuracy of $99\%$. We also developed a model-driven 3D positioning strategy which exploits the geometric relationships between the channel parameters and the positions of the BS and the user. This geometric localization strategy can operate in LOS and NLOS channels, providing sub-meter accuracies for $85\%$ of users in LOS channels and for $35\%$ of users in NLOS channels. We developed \textit{ChanFormer}, a location refinement network that enhances channel representation and identifies the most probable vehicle location.  After position refinement with \textit{Chanformer},  $95\%$ of users in the LOS channels achieve the localization accuracy of $0.58$ m, and $50\%$ of users in the NLOS channels achieve an accuracy of $0.77$ m. Overall, the performance has been improved by $59\%\sim 69\%$ depending on the scenario. The results demonstrate that the idea of attention is well-suited to the joint localization and communication problem, and also shows the potential of migrating advanced DNN architectures to the field of wireless communications. 

%Current work is for initial access, where the NLOS-only situation could be a bit concerning regarding the achieved accuracy. Integrating other onboard sensors could be a solution for higher accuracy. In addition, this research will be extended to vehicle location tracking using mmWave RX signals or sensor fusion. With historical channel and location information, solutions for occasional blockages in a trajectory could be provided. Attention based methods could also be used, which could outperform traditional network architectures for time series processing, e.g., \ac{LSTM} \cite{van2020review}, which processes the inputs sequentially and can be still prone to losing important earlier information due to memory limitations. Furthermore, in highly dynamic scenarios, an alternative approach to consider is \ac{V2V} sidelink assisted localization, which facilitates direct communication between vehicles without relying on connectivity through cellular infrastructures.

\bibliographystyle{IEEEtran}
\bibliography{refs}

%%%%%%%%%%%%%%%%%%%%%%%%%%%%%%%%%%%%%%%%%%%%%%%%%%%%%%%%%%%%%%%%%%%%%%%%%%%%%%%%%%%%%%%%%%%%%%%%%%%%%%%%%%%%%
\end{document}